\documentclass{aa}  
\usepackage{txfonts}
\usepackage{hyperref}

\usepackage{graphicx}	% Including figure files
\usepackage{amsmath}	% Advanced maths commands
\usepackage{amssymb}	% Extra maths symbols
\usepackage{xspace,bm}

\newcommand{\chandra}{\textit{Chandra}\xspace}

\newcommand{\fire}{\textit{Magellan}/FIRE\xspace}
\newcommand{\alma}{\textit{ALMA}\xspace}
\newcommand{\ang}{~\mathrm{\AA}\xspace}

\newcommand{\um}{~\mathrm{\mu m}\xspace}
\begin{document} 
\title{\chandra and \textit{Magellan/FIRE} follow-up observations of PSO167--13: an X-ray weak QSO at $z=6.515$}
   %   \subtitle{}
\titlerunning{The first heavily obscured QSO at $z>6$}
\authorrunning{F. Vito et al.}
   \author{F. Vito\thanks{fvito.astro@gmail.com}\inst{1,2} \and
          W. N. Brandt\inst{3,4,5} \and
          F. Ricci\inst{6,7}\and
          E. Congiu\inst{8} \and
          T. Connor\inst{9,10}\and
          E. Ba\~nados\inst{11}\and
          F. E. Bauer\inst{2,12,13} \and
          R. Gilli\inst{7} \and
          B. Luo\inst{14,15,16}\and
          C. Mazzucchelli\inst{17} \and
          M. Mignoli\inst{7} \and
          O. Shemmer\inst{18} \and
          C. Vignali\inst{6,7} \and
          F. Calura\inst{7}\and
          A. Comastri\inst{7} \and
          R. Decarli\inst{7} \and
          S. Gallerani\inst{1} \and
          R. Nanni\inst{19} \and
          M. Brusa\inst{6,7}\and
          N. Cappelluti\inst{20}\and
          F. Civano\inst{21}\and
          G. Zamorani\inst{7}
          }
 \institute{Scuola Normale Superiore, Piazza dei Cavalieri 7, 56126, Pisa (Italy)
        \and
Instituto de Astrof\'sica and Centro de Astroingenieria, Facultad de F\'sica, Pontificia Universidad Cat\'olica de Chile, Casilla 306, Santiago 22, Chile
\and
Department of Astronomy \& Astrophysics, 525 Davey Lab, The Pennsylvania State University, University Park, PA 16802, USA
\and
Institute for Gravitation and the Cosmos, The Pennsylvania State University, University Park, PA 16802, USA
\and
Department of Physics, The Pennsylvania State University, University Park, PA 16802, USA
\and
Dipartimento di Fisica e Astronomia, Universit\`a degli Studi di Bologna, via Gobetti 93/2, I-40129 Bologna, Italy
\and
INAF -- Osservatorio di Astrofisica e Scienza dello Spazio di Bologna, Via Gobetti 93/3, I-40129 Bologna, Italy
\and
Departamento de Astronom\'ia, Universidad de Chile, Camino del Observatorio 1515, Las Condes, Santiago, Chile
\and
Jet Propulsion Laboratory, California Institute of Technology, 4800 Oak Grove Drive, Pasadena, CA 91109, USA
\and
The Observatories of the Carnegie Institution for Science, 813 Santa Barbara St., Pasadena, CA 91101, USA
\and
Max Planck Institut f\"ur Astronomie, K\"onigstuhl 17, D-69117, Heidelberg, Germany
\and
Millennium Institute of Astrophysics (MAS), Nuncio Monse\~nor S\'otero Sanz 100, Providencia, Santiago, Chile
\and
Space Science Institute, 4750 Walnut Street, Suite 205, Boulder, Colorado, 80301, USA
\and
School of Astronomy and Space Science, Nanjing University, Nanjing 210093, PR China
\and
Key Laboratory of Modern Astronomy and Astrophysics, Nanjing University, Ministry of Education, Nanjing, Jiangsu 210093, PR China
\and
Collaborative Innovation Center of Modern Astronomy and Space Exploration, Nanjing 210093, PR China
\and
European Southern Observatory, Alonso de C\'ordova 3107, Vitacura, Regi\'on Metropolitana, Chile
\and
Department of Physics, University of North Texas, Denton, TX 76203, USA
\and
Department of Physics, University of California, Santa Barbara, CA 93106-9530, USA
\and
Physics Department, University of Miami, Coral Gables, FL 33124, USA
\and
Center for Astrophysics | Harvard \& Smithsonian, 60 Garden st, Cambridge, MA 02138, USA
%Millennium Institute of Astrophysics (MAS), Nuncio Monse\~nor S\'otero Sanz 100, Providencia, Santiago, Chile
%\and
%Space Science Institute, 4750 Walnut Street, Suite 205, Boulder, Colorado, 80301, USA
%\and
%INAF -- Osservatorio di Astrofisica e Scienza dello Spazio di Bologna, Via Gobetti 93/3, I-40129 Bologna, Italy
%\and
%School of Astronomy and Space Science, Nanjing University, Nanjing 210093, China
%\and
%Key Laboratory of Modern Astronomy and Astrophysics, Nanjing University, Ministry of Education, Nanjing, Jiangsu 210093, China
%\and
%Collaborative Innovation Center of Modern Astronomy and Space Exploration, Nanjing 210093, China
%\and
%European Southern Observatory, Alonso de C\'ordova 3107, Vitacura, Regi\'on Metropolitana, Chile
%\and
%Department of Physics, University of North Texas, Denton, TX 76203, USA
%\and
%Physics Department, University of Miami, Coral Gables, FL 33124, USA
%\and
%15 Center for Astrophysics | Harvard \& Smithsonian, 60 Garden st, Cambridge, MA 02138, USA
%\and
%Sorbonne Universit\'es, UPMC Universit\'e Paris 06 et CNRS, UMR7095, Institut d'Astrophysique de Paris, 98bis boulevard Arago, \\F-75014, Paris, France
  }

 \date{}
% \abstract{}{}{}{}{} 
% 5 {} token are mandatory
\abstract
 %{} %leave it empty if necessary  
 {The discovery of hundreds of quasi-stellar objects (QSOs) in the first Gyr of the Universe powered by  already grown supermassive black holes (SMBHs)  challenges our knowledge of SMBH formation. In particular, investigations of $z>6$ QSOs  presenting notable properties can provide unique information on the physics of fast SMBH growth in the early universe. }
   {We present the results of follow-up observations of the $z=6.515$ radio-quiet QSO PSO167--13, which is interacting with a close companion galaxy. The PSO167--13 system has been recently proposed to host the first heavily obscured X-ray source at high redshift. The goals of these new observations are to confirm the existence of the X-ray source and to investigate the rest-frame UV properties of the QSO.}
  {We observed the PSO167--13 system with \textit{Chandra}/ACIS-S (177 ks), and  obtained new spectroscopic observations (7.2 h) with \fire.}
  {No significant X-ray emission is detected from the PSO167--13 system, suggesting that the obscured X-ray source previously tentatively detected was either due to a strong background fluctuation or is highly variable. The upper limit ($90\%$ confidence level) on the X-ray emission of \mbox{PSO167--13} ($L_{2-10\,\mathrm{keV}}<8.3\times10^{43}\,\mathrm{erg s^{-1}}$) is the lowest available for a $z>6$ QSO. The ratio between the X-ray and UV luminosity of $\alpha_{ox}<-1.95$ makes PSO167--13 a strong outlier from the $\alpha_{ox}-L_{UV}$ and $L_X-L_{\mathrm{bol}}$ relations. In particular, its \mbox{X-ray} emission is $>6$ times weaker than the expectation based on its UV luminosity. The new \fire spectrum of \mbox{PSO167--13} is strongly affected by the unfavorable sky conditions, but the tentatively detected C IV and Mg II emission lines appear strongly blueshifted. }
{The most plausible explanations for the X-ray weakness of PSO167--13 are intrinsic weakness or small-scale absorption by Compton-thick material. The possible strong blueshift of its emission lines hints at the presence of nuclear winds, which could be related to its X-ray weakness.}


\maketitle
%
%-------------------------------------------------------------------

\section{Introduction}

In the last two decades, $>200$ quasi-stellar objects (QSOs) have been discovered at $z>6$, when the Universe was $<1$ Gyr old, primarily thanks to the availability of wide-field optical/NIR surveys \citep[e.g.,][]{Banados16,Banados18a,Matsuoka18a,Fan19, Reed19, Belladitta20, Yang20, Wang21b}. 
The selection of high-redshift QSOs is based on the detection of the bright rest-frame UV nuclear continuum, with all of the currently confirmed z > 6 QSOs classified optically as type 1 (i.e., unobscured; but see, e.g., \citealt{Matsuoka19}).  Therefore, while theoretical arguments and numerical simulations (e.g., \citealt{Pacucci15,Valiante17}) usually require long periods of fast and heavily obscured mass growth onto black-hole seeds ($10^2-10^5\,M_\odot$; e.g., \citealt{Woods19} and references therein) in order to explain the presence of $1-10$ billion $M_\odot$ supermassive black holes (SMBHs) at $z>6$ \citep[e.g.,][]{Wu15}, very little is known about the population of obscured accreting SMBHs in the early universe from an observational point of view.

The accumulation of multiwavelength data for a continuously increasing number of optically selected high-redshift QSOs has recently made possible the first statistical studies of QSOs in the early universe.  These new data and analyses have improved our understanding of the mechanisms of SMBH formation and early growth, their interplay with their host galaxies, their environments, and the physics of reionization
 \citep[e.g.,][]{Decarli18, Davies19,Farina19,Mazzucchelli19, Neeleman19,Onoue19,  Eilers20, Schindler20, Wang21a}. One of the key results is that the observable spectral energy distribution properties of high-redshift QSOs do not appear to differ strongly from their counterparts at later cosmic times, in particular concerning the UV and X-ray emission, that trace the accretion physics close to the accreting SMBHs (e.g., \citealt{DeRosa14, Gallerani17, Mazzucchelli17b,Nanni17, Vito19b}). However, recent results do point toward a larger fraction of weak-line QSOs (WLQs; \citealt{Shen19}) and larger blueshifts of high-ionization UV emission lines at $z>6$ \citep{Meyer19,Schindler20}, suggesting a high incidence of nuclear winds in these systems.

In addition to statistical sample studies, a few high-redshift QSOs have been the targets of more focused investigations into their properties \citep[e.g.,][]{Eilers18, Connor19, Connor20,Fan19, Nanni18,Mignoli20, Spingola20, Wang21b}.
In \cite{Vito19a}, we discussed the peculiar \mbox{X-ray} properties of PSO J167.6415--13.4960 (hereafter \mbox{PSO167--13}; RA$_{\mathrm{ICRS}}$=11:10:33.963, DEC$_{\mathrm{ICRS}}$=$-$13:29:45.73; \citealt{Venemans15a,Venemans20}), a type 1 QSO at $z=6.515$ (systemic redshift derived from the [C II] 158 $\mu$m emission line; \citealt{Decarli18}). The UV luminosity of PSO167--13 ($M_{1450\ang}=-25.6$) places this QSO close to the break of the UV luminosity function of $z>6$ QSOs \citep[e.g.][]{Jiang16,Matsuoka18c}.
Atacama Large Millimeter/submillimeter Array (\alma) imaging revealed that PSO167--13 is interacting with a companion galaxy at a projected distance of $\approx0.9^{\prime\prime}$ (i.e., $\approx5$ physical kpc at $z=6.515$) and $\Delta v \approx -140\, \mathrm{km\,s^{-1}}$ (i.e., $\Delta z \approx-0.004$) in velocity space \citep{Willott17,Neeleman19}. Companion galaxies detected with \alma have been found for a significant fraction of high-redshift QSOs \citep{Decarli18, Venemans20}.
The PSO167--13 system is one of only two cases in which the companion galaxy has been detected in the rest-frame UV via deep \textit{Hubble} Space Telescope (\textit{HST}) imaging \citep{Decarli19,Mazzucchelli19}. 

 A 59 ks \chandra observation of PSO167--13 revealed no counts in the 0.5-2~keV band, and three counts in the 2--5~keV band \citep{Vito19b,Vito19a}. Due to the low background level and excellent ($\approx0.5^{\prime\prime}$) PSF of \chandra, the three counts represented a relatively significant (99.96\% confidence level) detection. The lack of a soft-band counterpart suggested that the X-ray source was heavily obscured. 
 Although the position of the X-ray source suggested an association with the companion galaxy (with a spatial offset of $\approx0.1^{\prime\prime}$, to be compared with an offset of $\approx1^{\prime\prime}$ from the QSO), due to the close separation of the QSO-companion system, and the positional uncertainty of the X-ray emission, we could not associate it unambiguously to one of the two galaxies.
 
Irrespective of the lack of a secure identification of the \mbox{X-ray} source with either of the two ALMA-detected galaxies, the optically selected QSO was not detected in the soft X-ray band, either because of obscuration or intrinsic faintness. \mbox{X-ray} weakness (i.e.,  X-ray emission fainter than the expectation based on known relations with the UV or bolometric luminosity) is often found in notable classes of optically classified type-1 QSOs, such as broad absorption line QSOs (BAL QSOs; e.g., \mbox{\citealt{Luo14}}, \citealt{Vito18c}), WLQs (e.g., \citealt{Luo15,Ni18}), and ``red'' QSOs \citep[e.g., ][]{Pu20}. The available rest-frame UV spectrum of PSO167--13 \citep[0.73h \textit{VLT/FORS2} using a $1.3^{\prime\prime}$ slit, $+$ 3.33h \textit{Magellan/FIRE}, using a $0.6^{\prime\prime}$ slit]{Venemans15a} does not allow us classify it securely as a BAL QSO or a WLQ. However, the C IV (1549$\ang$) emission line appears weak and blueshifted relative to the systemic redshift, as is often found for WLQs \citep[e.g., ][and references therein]{Ni18} and, in particular, $z>6$ QSOs (e.g., \citealt{Schindler20}). Moreover, the noisy spectrum blueward of the C IV and Si IV emission lines could hide possible absorption features, characteristic of BAL QSOs. With so many uncertainties in the nature of this system, deeper X-ray and rest-frame UV observations are needed.
 
 Here we present new \chandra X-ray observations and \textit{Magellan} Folded-port InfraRed
 Echellette (FIRE; \citealt{Simcoe08})  rest-frame UV spectroscopy of PSO167--13. The goals of these observations are to confirm the presence of an obscured X-ray source in the PSO167--13 system, constrain the level of X-ray weakness of the QSO, and investigate its possible BALQSO or WLQ nature. Errors are reported at 68\% confidence levels, while limits are given at 90\% confidence levels, unless otherwise noted. We adopt a flat cosmology with $H_0=67.7\,\mathrm{km\,s^{-1}}$ and $\Omega_m=0.307$ \citep{Planck16}. At $z=6.515$, $1^{\prime\prime}$ corresponds to a physical distance of 5.6 kpc.

\begin{figure*}
	\begin{center}
		\hbox{
			\includegraphics[width=180mm,keepaspectratio]{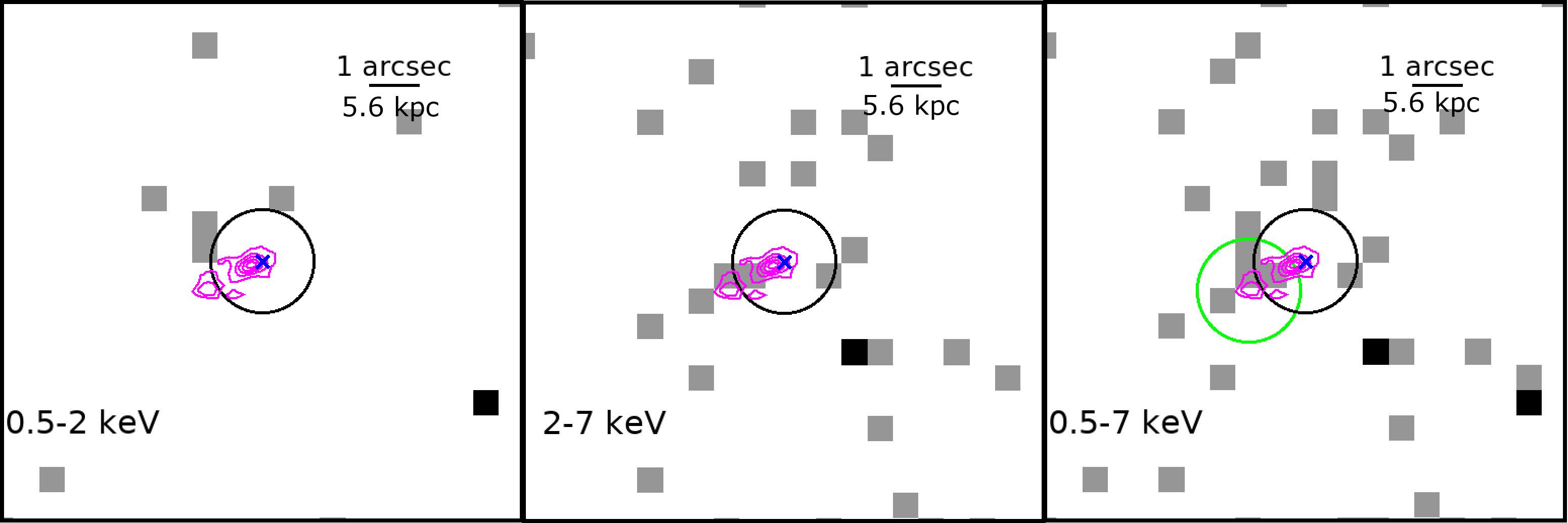} 
		}
	\end{center}
	\caption{ \chandra $5^{\prime\prime}\times5^{\prime\prime}$ images in the soft, hard, and full bands for the total 177 ks exposure. North is up, East is left. The magenta contours show the [C II] emission detected with ALMA ($0.33^{\prime\prime}\times0.22^{\prime\prime}$ beam size) of the PSO167--13 (main emission) and companion (fainter emission to the south-east) system from $4\sigma$ to $20\sigma$ in steps of $4\sigma$ (\citealt{Venemans20}, see also \citealt{Neeleman19}). The blue X symbols mark the optical position of PSO167--13 \citep{Venemans20}. The black circles denote the $1^{\prime\prime}$ extraction radius used throughout the text for PSO167--13. The green circle indicates the $1^{\prime\prime}$ extraction radius used for the companion galaxy (shown only in the right panel for clarity).}\label{Fig_Xray_images}
\end{figure*}

\section{Observations and data analysis}
In this section we briefly describe the \chandra and \fire observations of PSO167--13, and their reduction.

\subsection{\chandra observations}\label{Chandra_obs}
In Cycle 21, we observed PSO167--13 with \chandra ACIS-S (S3 chip) for 117.4~ks. Considering also the 59.3 ks dataset taken in 2018 \citep{Vito19a}, the total \chandra coverage of PSO167--13 is $\approx177$ ks (see Tab.~\ref{Tab_X-ray_obs} for a summary of the \mbox{X-ray} pointings). We reprocessed the \textit{Chandra} observations with the \textit{chandra\_repro} script in CIAO 4.12  \citep{Fruscione06},\footnote{http://cxc.harvard.edu/ciao/} using CALDB v4.9.3,\footnote{http://cxc.harvard.edu/caldb/} setting the option \textit{check\_vf\_pha=yes} in the case of observations taken in Very Faint mode.

In order to correct the astrometry of each observation, we performed source detection with the \textit{wavdetect} script with a no-source probability threshold of $10^{-6}$ on each individual \chandra exposure. Then, we used the \textit{wcs\_match} and \textit{wcs\_update} tools to match the positions of the X-ray sources with $>10$ counts  to objects in the Pan-STARRS DR2 source mean catalog \citep{Chambers16}\footnote{https://panstarrs.stsci.edu/}  and correct the astrometry of the X-ray observations. The astrometry of the Pan-STARRS DR2 catalog is in turn registered to the Gaia DR2 astrometry \citep{GAIADR2}. We could not directly use the Gaia catalog due to the small number of reliable counterparts of the X-ray sources.
Finally, we merged the individual observations with the \textit{reproject\_obs} tool, and derived merged images and exposure maps. We repeated the detection procedure on the merged observation, and found a median difference between the position of the X-ray sources and the Pan-STARRS catalog of $\approx0.3^{\prime\prime}$, which is consistent with the \chandra pixel size ($\approx0.5^{\prime\prime}$). Only a few Gaia entries are associated unambiguously with X-ray sources, with offsets of $\approx$0.3--0.4$^{\prime\prime}$, providing a useful consistency check.

 We extracted response matrices and ancillary files from individual pointings using the \textit{specextract} tool and added them using the \textit{mathpha}, \textit{addrmf}, and \textit{addarf} HEASOFT tools\footnote{\url{https://heasarc.gsfc.nasa.gov/docs/software/heasoft/}}, weighting by the individual exposure times. Ancillary files, which are used to derive fluxes and luminosities, were aperture corrected.

\begin{table}
	\caption{Summary of the \chandra\, observations of PSO167--13. Data from OBSID 20397 were already presented in \citet{Vito19b,Vito19a}.}
	\begin{tabular}{ccccccccc} 
		\hline
		\multicolumn{1}{c}{{ OBSID}} &
		\multicolumn{1}{c}{{ Start date }} &
		\multicolumn{1}{c}{{ $T_{exp}$ [ks]}} \\ 
		20397 &2018-02-20 &59.3\\
		22523 &2020-02-12 &42.8\\
		23153 &2020-02-15 &24.0\\
		23018 &2020-03-28 &10.0\\
		23199 &2020-03-29 &40.6\\
		\hline
	\end{tabular} \label{Tab_X-ray_obs}\\
\end{table}

\subsection{\fire observations}

We obtained near-infrared (NIR) spectroscopy for PSO167--13 in the range $\lambda=0.8-2.5\,\mu$m with \fire during three nights (March 3rd-5th 2020), for a total of 9.9 h on source. Observations were conducted during gray time, using the $1^{\prime\prime}$ wide slit in the high-resolution echellete mode, with a nominal resolution of R=3600.
During these observations, the sky conditions were unstable, with rapidly varying seeing ($\approx0.7^{\prime\prime}-2^{\prime\prime}$) and sky background. We rejected the exposures with seeing $>1.5^{\prime\prime}$, lowering the useful amount of on-source time to 7.2 h on source. The individual spectra were obtained using the nodding technique ($3^{\prime\prime}$ nod length) in a sequence of ABBA acquisitions (44 exposures, each of 602.4 s). The data were reduced with the Interactive Data Language (IDL) pipeline FireHose v2 package \citep{Gagne15}, and custom Python scripts. OH airglow was used %followed by short  arc frames (ThAr) in order 
to correct  for telescope flexure and obtain the wavelength solution. Nearby A0V stars with airmass similar to that of the target were observed after each ABBA block in order to derive telluric absorption corrections and absolute flux calibrations, which we applied to the corresponding ABBA block, and define the extraction traces. We corrected for Galactic extinction ($E(B-V)=0.0485$; from \cite{Schlafly11}) with the extinction curve of \cite{Fitzpatrick99}.

% \subsection{\xshooter observations}
% PSO167--13 was observed for 40 min on source with \xshooter in 2017 (Program ID 097.B-1070(A), PI: Farina). We retrieved the data from the ESO archive and reduced them with the standard pipeline.\footnote{\url{https://www.eso.org/sci/software/esoreflex/}} According to the log file, the seeing during the observations was $\approx0.8^{\prime\prime}$.
% 
% %Visual inspection of the VIS arm of \xshooter revealed possible emission lines blueward of the PSO167--13 Lyman break (at rest-frame $912\ang$), thus not consistent with being emitted from the QSO. If confirmed with deeper spectroscopy, these lines will pinpoint the presence of a foreground galaxy in the direction of PSO167--13.
% We display the 2-dimensional spectrum at the wavelenghts of such lines in Appendix~\ref{AppendixA}.

 \section{Results}
 In this section, we report the results derived from the \chandra observations of PSO167--13 and the companion galaxy, and \fire spectroscopic observations of PSO167--13.

\subsection{X-ray photometry of the companion galaxy}\label{galaxy}

We present in Fig.~\ref{Fig_Xray_images} the \chandra images in the soft (0.5--2 keV), hard (2--7 keV), and full (0.5--7 keV) bands of the \mbox{PSO167--13} system.
%In \cite{Vito19a} we detected three counts in the $2-5$ keV band consistent with the position of the companion galaxy, although, given the positional uncertainty, we could not exclude that the emission was due to the QSO.
As we did in \cite{Vito19a}, here we use a $1^{\prime\prime}$ radius extraction region (green circle in Fig.~\ref{Fig_Xray_images}; corresponding to $\approx90\%$ encircled energy fraction at 1.5 keV), centered on the ALMA position of the companion galaxy (RA=11:10:34.033, DEC=$-$13:29:46.29; \citealt{Neeleman19}) to compute its photometry. We computed the detection significance in each energy band using the binomial no-source probability $P_{B}$ presented by \cite{Weisskopf07} and \cite{Broos07}, and set a significance threshold $(1-P_B)=0.99$ for source detection. We measured the X-ray background in an annular region with $R_{in}=4^{\prime\prime}$ and $R_{out}=24^{\prime\prime}$, where no bright X-ray sources are found. We detect 1, 3, and 4 counts in the soft, hard, and full bands. Considering the expected background of $0.35/0.59/0.93$ counts, respectively, we obtained a detection significance $(1-P_B)<0.99$ (i.e., the galaxy is not detected) in all of the bands. Following the method of \cite{Weisskopf07}, we computed upper limits on the net counts $<3.6$, $<6.1$, and $<7.1$ in the soft, hard, and full bands, respectively. 

 In order to compute upper limits on the observed flux in the three bands, we assumed power-law emission with $\Gamma=2$. This $\Gamma$ value is typical of rapidly accreting SMBHs \citep[e.g.,]{Shemmer08,Brightman13} and is consistent with the average photon index derived for optically selected QSOs up to $z\approx7.5$ \citep[e.g.,][]{Nanni17,Banados18b,Vito18b}, although hints for steeper photon indexes at $z\gtrsim6.5$ have been reported by 
 	\cite{Vito19b} and 
 	\mbox{\cite{Wang21a}}. 
 Accounting also for Galactic absorption toward the PSO167--13 system ($N_H=4.7\times 10^{20}\,\mathrm{cm^{-2}}$; e.g., \citealt{Kalberla05}), we estimate fluxes $F<(1.9/6.6/5.0)\times10^{-16}\,\mathrm{erg\,cm^{-2}\,s^{-1}}$ in the soft/hard/full bands. Although the exact values of the upper limits depend on the choice of $\Gamma$, the errors on the flux are largely dominated by the statistical uncertainties on the X-ray counts. We computed the upper limit on the intrinsic 2--10 keV luminosity from the soft-band flux, consistently with previous works \citep[e.g.;][]{Vito19a,Vito19b}, as $L_{X}<1.3\times10^{44}\,\mathrm{erg\,s^{-1}}$.

 \begin{figure}
 	\begin{center}
 		\hbox{
 			\includegraphics[width=90mm,keepaspectratio]{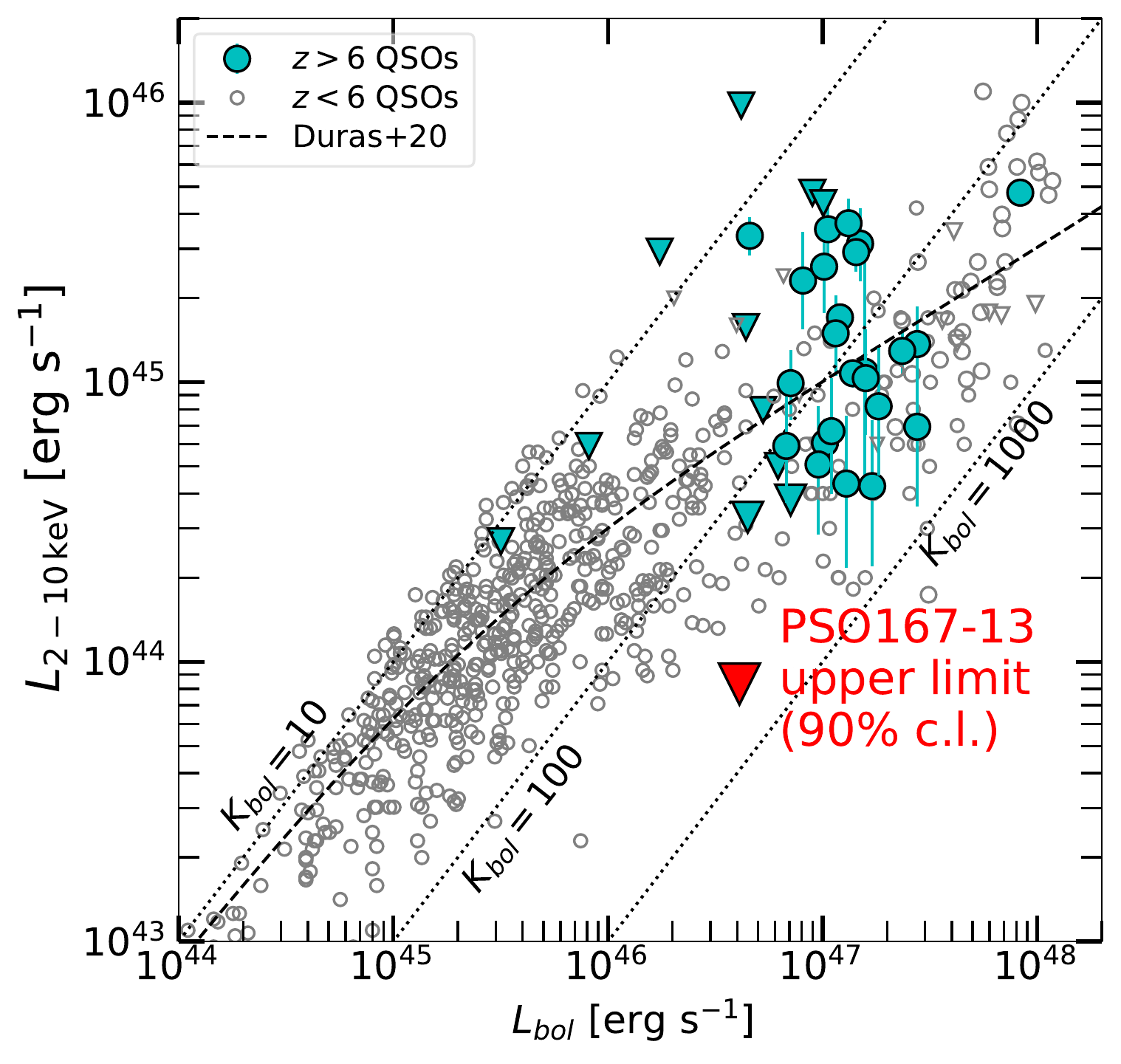} 
 		}
 	\end{center}
 	\caption{ X-ray luminosity versus bolometric luminosity for $z>6$ QSOs (cyan symbols) from \citet{Connor19,Connor20}, \citet{Vito19b}, \citet{Pons20}, and \citet{Wang21a}, compared with $z<6$ QSOs (grey symbols) from \citealt{Lusso12, Martocchia, Nanni17, Salvestrini19}, and the $L_X-L_{bol}$ relation (black dashed line) from \citet{Duras20}. Only radio-quiet QSOs are shown. Circles represent detected sources, and downward-pointing triangles mark upper limits. PSO167--13 is shown in red. Diagonal dotted lines mark the loci of constant bolometric correction (i.e., $K_{bol}=L_{bol}/L_{X}$).}\label{Fig_Lx_Lbol}
 \end{figure}
 
 \subsection{X-ray photometry of PSO167--13}\label{PSO167}
We computed the X-ray photometry of PSO167--13 in a circular region of $R=1^{\prime\prime}$ centered on the  optical/NIR position of the QSO (RA=11:10:33.638, DEC=$-$13:29:45.73), which is provided by \cite{Venemans20} based on the Gaia DR2 astrometry. They also report a small offset ($\approx0.15^{\prime\prime}$) between the optical/NIR coordinates (blue cross in Fig.~\ref{Fig_Xray_images}) and the peak position of the dust continuum and [C II] emission line of the QSO (magenta contours), possibly due to the QSO host galaxy being stretched during the ongoing interaction with the close companion galaxy \citep{Venemans20} and to the presence of a gas ``bridge" component \citep{Neeleman19}.

We detected 0, 2, and 2 counts in the soft, hard, and full bands, corresponding to $<2.3$, \mbox{$<4.8$}, and \mbox{$<4.5$} net counts\footnote{The slightly lower upper limit on the net counts in the full band than that in the hard band is due to the higher background level.}, respectively \citep{Weisskopf07}.  
As in \S~\ref{galaxy}, we converted the upper limits on the counts into fluxes of \mbox{$F<(1.2/5.1/3.2)\times10^{-16}\,\mathrm{erg\,cm^{-2}\,s^{-1}}$} in the soft/hard/full bands and a rest-frame 2-10 keV luminosity of \mbox{$L_{X}<8.3\times10^{43}\mathrm{erg\,s^{-1}}$}. We note that one count in the hard band lies within the extraction regions of both the QSO and the companion galaxy, due to their small angular separation, such that the fluxes in the hard and full bands for at least one of these galaxies are overestimated.
Fig.~\ref{Fig_Lx_Lbol} presents the X-ray luminosity versus bolometric luminosity\footnote{Bolometric luminosities for $z>6$ QSOs and for the \cite{Nanni17} and \cite{Salvestrini19} samples are estimated from $M_{1450\ang}$, using the bolometric correction of \citet[see also \citealt{Decarli18}]{Venemans16}, while values derived from SED fitting are plotted for the QSOs in the  \cite{Lusso12} and \cite{Martocchia} samples.}
 of QSOs at $z>6$ and lower redshift. The upper limit on PSO167--13 at $L_{bol}=4.1\times10^{46}\,\mathrm{erg\,s^{-1}}$ is significantly lower than the X-ray luminosity of X-ray detected QSOs at $z>6$, and is a stronger constraint than the available upper limits on other undetected sources. The upper limit on $L_X$ for PSO167--13 translates into a bolometric correction $K_{bol}=L_{bol}/L_{X}>492$, to be compared with a typical value of $K_{bol}\approx100$ for QSOs with similar bolometric luminosities.

\begin{figure}
	\begin{center}
		\hbox{
			\includegraphics[width=90mm,keepaspectratio]{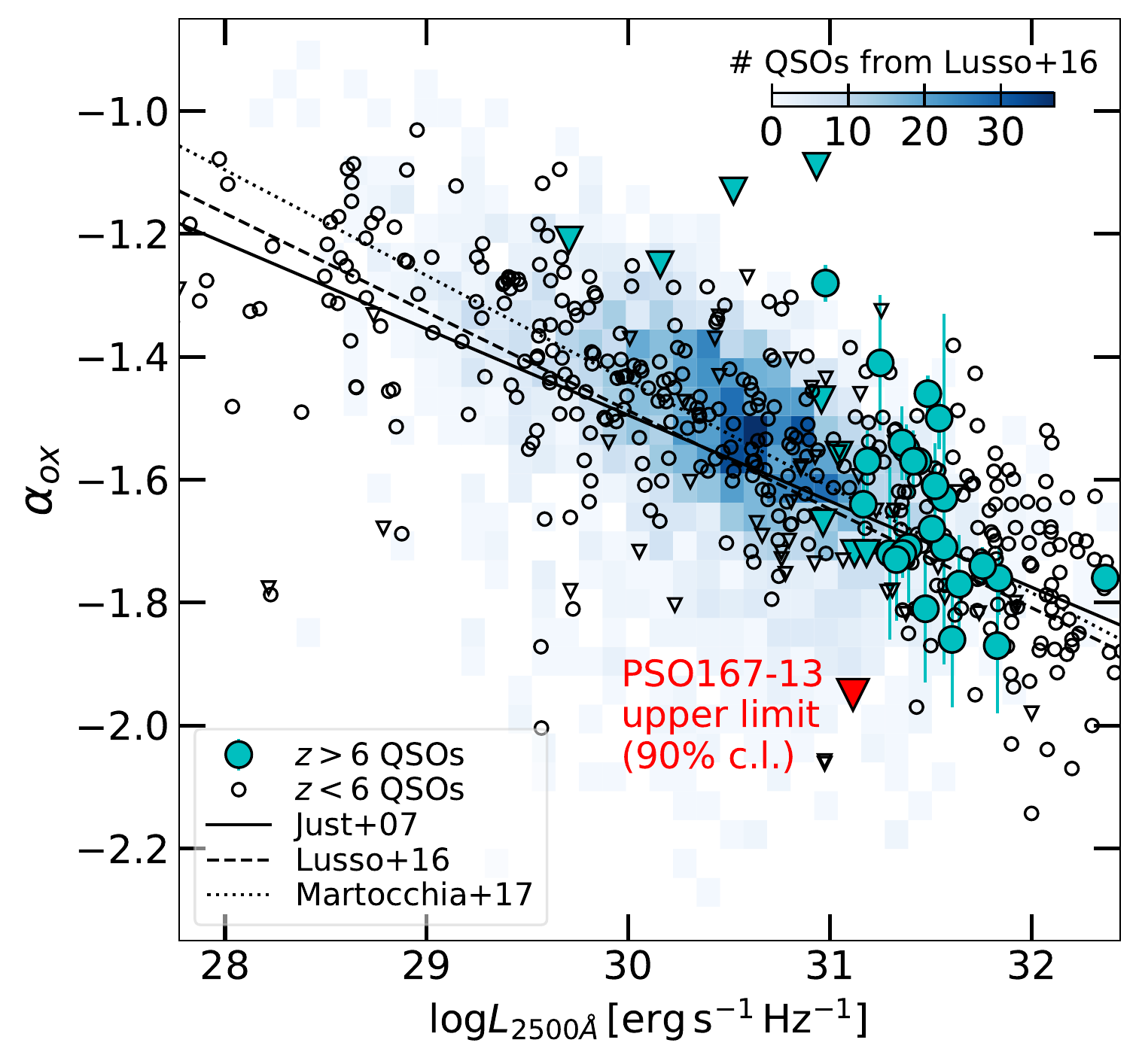} 
		}
	\end{center}
	\caption{ $\alpha_{ox}$ plotted against UV luminosity for PSO167--13 (in red) and other $z>6$ QSOs (cyan filled symbols) from \citet{Connor19,Connor20}, \citet{Vito19b}, \citet{Wang21a}, and the updated values of the \citet{Pons20} $z>6.5$ QSOs (\citealt{Pons21}). We also show $z<6$ QSOs (black empty symbols and blue color map) from \citet{Shemmer06, Steffen06, Just07, Lusso16, Nanni17, Salvestrini19}, and best-fitting relations from \citet{Just07}, \citet{Lusso16}, \citet{Martocchia}. Circles represent detected sources, downward-pointing triangles mark upper limits. For visual purposes, we do not plot X-ray undetected sources included in the \citet{Lusso16} sample. }\label{Fig_aox_L2500}
\end{figure}

The relative contribution of the X-ray and UV emission in QSOs is usually parametrized by the quantity \hbox{$\alpha_{ox}=0.38\times \mathrm{log}(L_{2\,\mathrm{keV}}/L_{2500\ang})$}, which represents the slope of a nominal power-law connecting the emission in the two bands (e.g., \citealt{Brandt15} and references therein). We measure $L_{2500\ang}=1.3\times10^{31}\,\mathrm{erg\,s^{-1}Hz^{-1}}$ from the best-fitting UV continuum of the 2020 FIRE spectrum of PSO167--13 (see \S~\ref{FIRE}) and convert the upper-limit on $L_{2-10\,\mathrm{keV}}$ into \mbox{$L_{2\,\mathrm{keV}}<1.07\times10^{26}\,\mathrm{erg\,s^{-1}\,Hz^{-1}}$} (assuming power law emission with $\Gamma=2$). From these values, we derived $\alpha_{ox}<-1.95$, which is the lowest value for a $z>6$ QSO, and among the lowest values for the general QSO population (see Fig.~\ref{Fig_aox_L2500}).

\begin{figure}
	\begin{center}
		\hbox{
			\includegraphics[width=90mm,keepaspectratio]{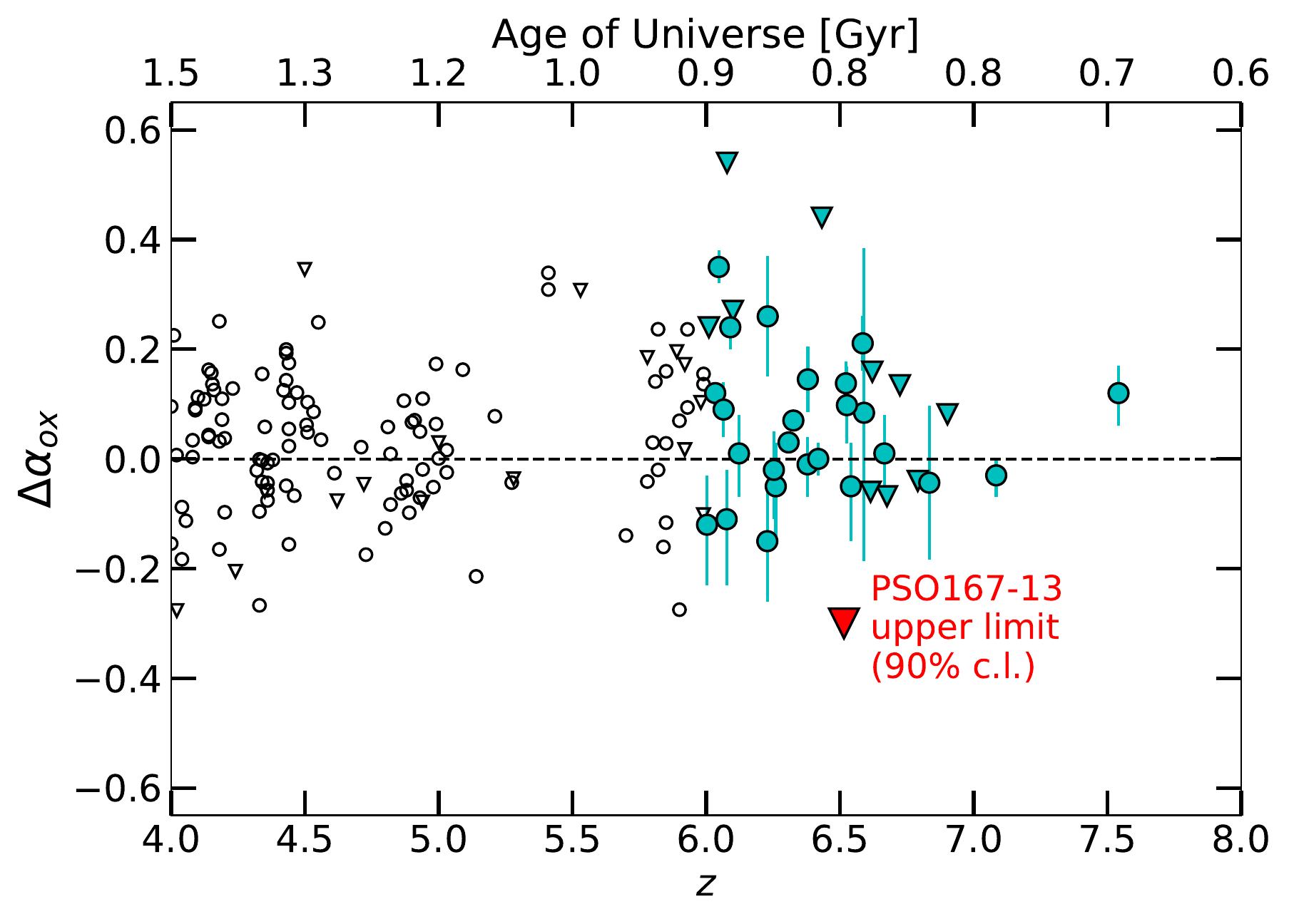} 
		}
	\end{center}
	\caption{ $\Delta\alpha_{ox}$ as a function of redshift for $z>4$ QSOs. Symbols are the same as in Fig.~\ref{Fig_aox_L2500}. The upper limit on PSO167--13 suggests it is the most extreme X-ray weak quasar yet known beyond $z>4$.}\label{Fig_daox_z}
\end{figure}

A well-known anti-correlation exists between $\alpha_{ox}$ and $L_{\mathrm{UV}}$ up to $z>6$ \citep[e.g.,][]{Just07,Lusso16,Martocchia,Nanni17,Vito19b}. Therefore, a fairer comparison between QSOs with different UV luminosities can be made considering the values of $\Delta\alpha_{ox}=\alpha_{ox}^{obs} - \alpha_{ox}^{exp}$; i.e., the difference between the observed $\alpha_{ox}$ and the value expected for a given QSO's UV luminosity. Assuming the \cite{Just07} relation, as in \cite{Vito19b}, we find for \mbox{PSO167--13} $\Delta\alpha_{ox}<-0.30$. The value of $\Delta\alpha_{ox}$ for PSO167--13 implies a  factor of $\gtrsim6$ weaker X-ray emission than the expectation, in agreement with the QSO location in Fig.~\ref{Fig_Lx_Lbol}. Among X-ray detected QSOs at $z>6$, none shows such a level of X-ray weakness, and undetected sources have shallower upper limits (Fig.~\ref{Fig_daox_z}).

\begin{figure*}
	\begin{center}
		\hbox{
			\includegraphics[width=180mm,keepaspectratio]{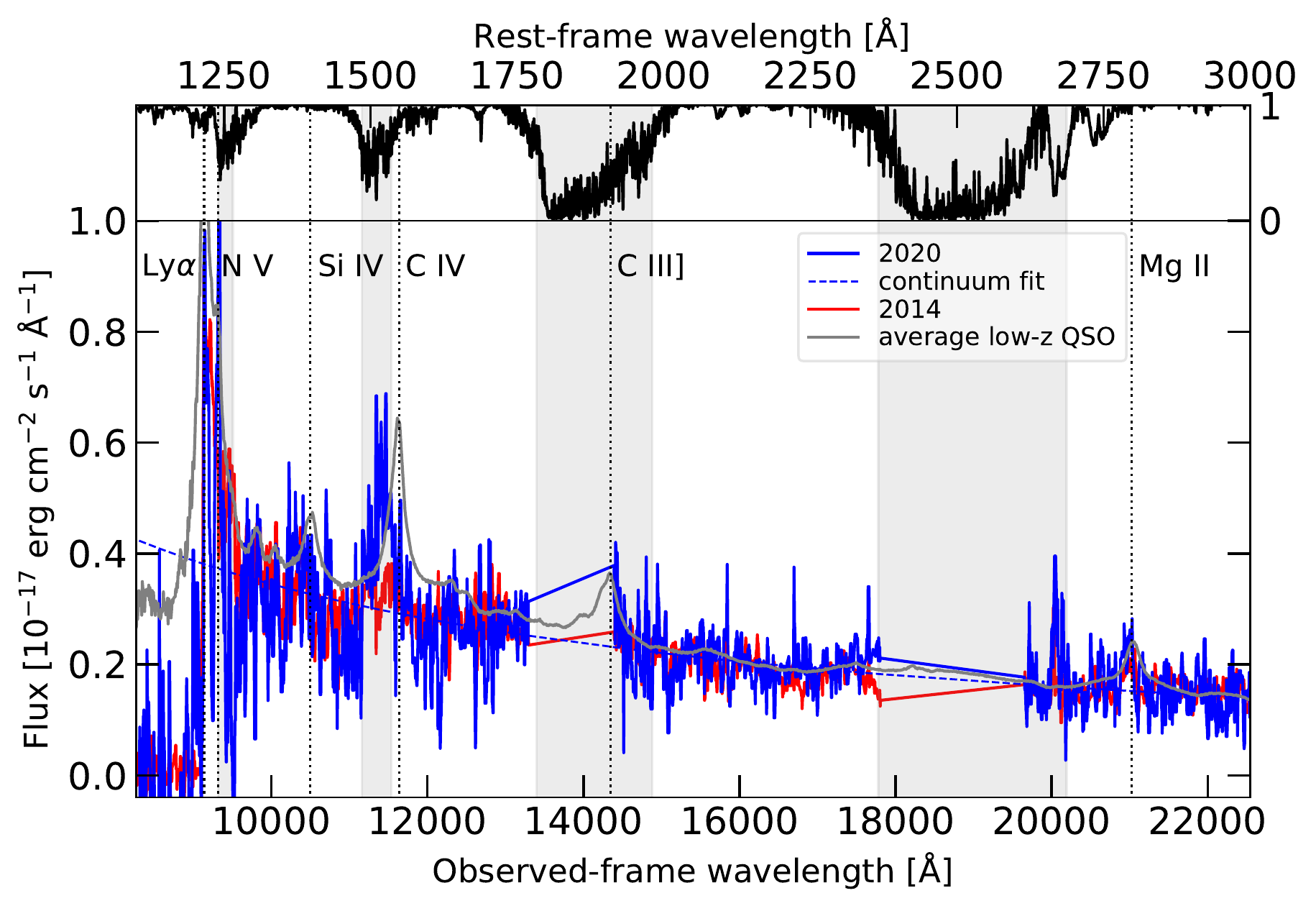} 
		}
	\end{center}
	\caption{ \textit{Bottom panel}: rest-frame UV spectrum of PSO167--13 obtained with our 2020 \fire observations (blue solid line), compared with the \citet{Venemans15a} spectrum (taken in 2014; red solid line). %We applied a Gaussian smoothing with $\approx25\ang$ kernel.
		 We applied a median filter with $\approx20\,\ang$ kernel for visual purpose, as in \citet{Mazzucchelli17b}.
		The average low-redshift QSO spectrum of \citet{VandenBerk01}, normalized to the rest-frame $3000\ang$ flux of the \citet{Venemans15a} spectrum, is shown as a grey line. We also present the best-fitting UV power-law continuum of the 2020 spectrum as a dashed blue line, and the expected location of QSO emission lines at $z=6.515$ as vertical dotted lines. \textit{Top panel:} example of atmospheric transmission during our FIRE observations. Regions with transmission $<0.6$ are marked with grey stripes in both panels. In particular, two spectral windows centered at rest-frame $\approx1800\ang$ and $\approx2500\ang$ are completely affected by the very low atmospheric transmission, and are thus masked in the spectra. }\label{Fig_FIRE}
\end{figure*}

\subsection{Rest-frame UV spectroscopy of PSO167--13}\label{FIRE}

Fig.~\ref{Fig_FIRE} presents the rest-frame UV spectrum of PSO167--13 we obtained with \fire in 2020. We compare it with the 2014 spectrum presented by \cite{Venemans15a}. 
 After flux calibration, the normalization of the 2020 spectrum is $\approx15\%$ lower than that reported in \cite{Venemans15a}, most likely due to the	
	varying seeing between the target and standard-star observations affecting the flux calibration in 2020. We therefore normalized the 2020 spectrum to the 2014 flux at 3000$\ang$.
Strong atmospheric absorption completely suppresses the QSO emission at rest-frame $\approx1790-1900\ang$ and $\approx2360-2690\ang$ (see the upper panel in Fig.~\ref{Fig_FIRE}). Therefore, we masked these two spectral windows in Fig.~\ref{Fig_FIRE}.

We fitted the rest-frame UV continuum to the unbinned 2020 spectrum in the spectral regions 2000-2350 $\ang$, 2690-2750 $\ang$, and 2850-3000 $\ang$ assuming a power-law of the form

\begin{equation}\label{UV_cont}
F_\lambda=F_0\left(\frac{\lambda}{2500\ang}\right)^\alpha.
\end{equation}
The best-fitting slope $\alpha=-1.10\pm0.12$ is in agreement with the result of \cite{Mazzucchelli17b}, and is redder than the typical QSO value ($\alpha=-1.7$; e.g., \citealt{Selsing16}, see also \citealt{Venemans15a}).

\subsubsection{Caveats on the rest-frame UV spectrum}

Despite the longer on-source exposure, the 2020 \fire spectrum is noisier than the 2014 spectrum of PSO167--13, due to the poor atmospheric conditions reported in \S~\ref{FIRE}. For instance, in the $H$ band (where no strong QSO emission lines are expected at $z=6.515$), we estimate a signal-to-noise ratio (SNR)\footnote{We used the DER\_STEN algorithm of \cite{Stoehr08}, available at \url{http://www.stecf.org/software/ASTROsoft/DER_SNR/}} of $\approx3.0$ and $\approx4.5$ for the 2020 and the 2014 spectra, respectively. In the following subsections, we describe the main spectral features visible in the spectrum, and the parameters derived with a basic analysis, for completeness. However, we warn that the results should be treated as merely indicative and with caution. 

\subsubsection{C IV emission line}
\begin{figure}
	\begin{center}
		\hbox{
			\includegraphics[width=83mm,keepaspectratio]{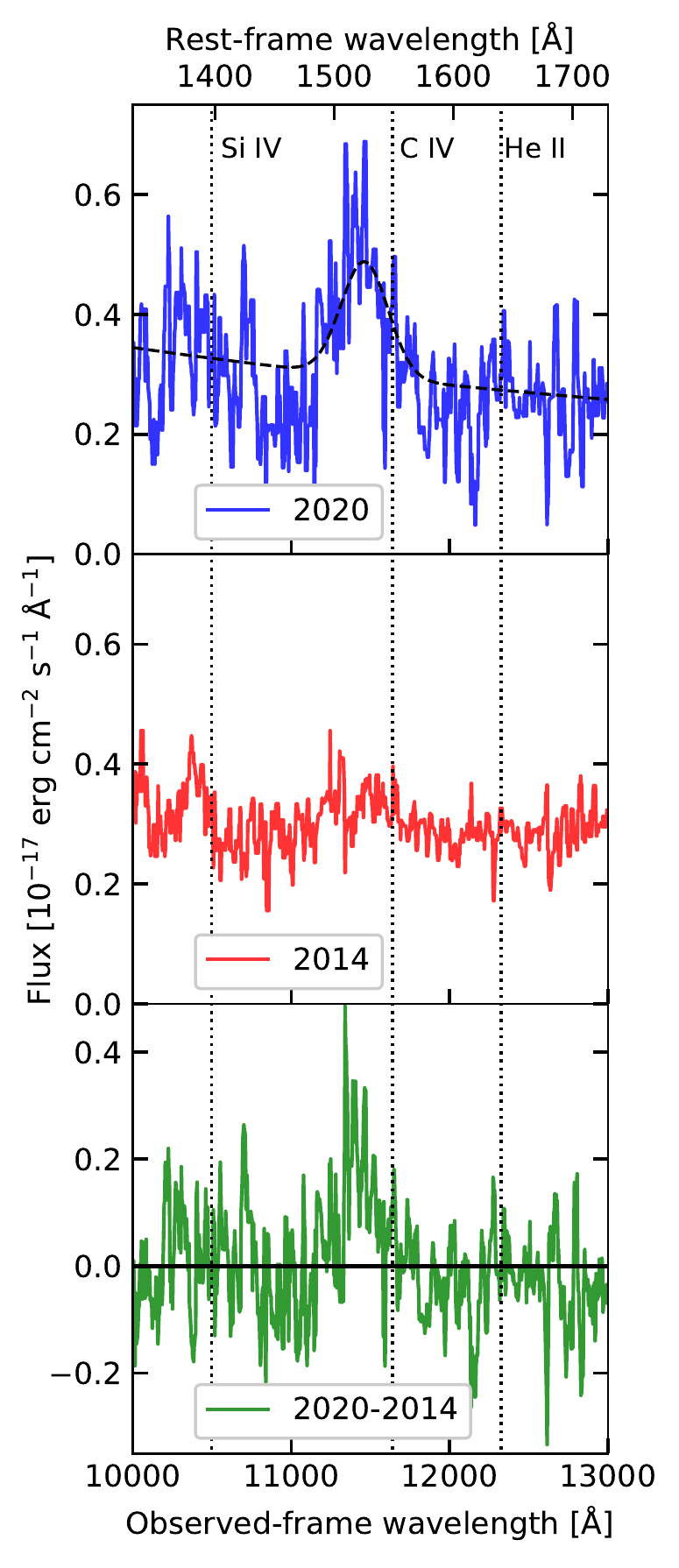} 
		}
	\end{center}
	\caption{ Zoom of the spectral region of the 2020 (top panel) and 2014 (middle panel) spectra encompassing the C IV emission line of PSO167--13. The bottom panel presents the difference of the two epochs. The dashed black line in the top panel marks the best-fitting continuum plus single Gaussian model of the C IV emission line.%, but we also add a short (20 min) \xshooter observation taken in 2017 (green line).
		% 	 The cyan regions mark the possible broad absorption lines discussed in the text. 		
	}\label{Fig_CIV}
\end{figure}

Fig.~\ref{Fig_CIV} zooms into the  $1.0-1.3\,\mathrm\um$ spectral region. %The $Ly\alpha$ emission line is clearly detected in both the 2014 and 2020 spectra, although its bluer wing is suppressed  by the $Ly\alpha$ forest. The Si IV (1397 $\ang$)  emission line appears weak and possibly blueshifted. We note that in this spectral range the 2020 spectrum appears noisier than the 2014 spectrum. The latter includes exposures taken with \textit{VLT}/FORS2, which is more efficient than \textit{Magellan}/FIRE at these observed wavelengths, although we cannot exclude an intrinsic variation of the spectrum. 
The C IV  (1549$\ang$) emission line was not clearly detected in the 2014 spectrum \citep{Mazzucchelli17b}, but there is a tentative detection of this emission line in the 2020 spectrum, with a nominal total $\mathrm{SNR} \approx5$.  Since it falls in a spectral region of relatively strong telluric absorption, it is unclear whether such a line is real or is partially or totally an artifact due to atmospheric correction. In this respect, future observations of PSO167--13 with \textit{JWST} can confirm the presence and properties of this line.

Assuming this feature is real, we fitted the unbinned spectrum with a single Gaussian function to derive basic parameters using a $\chi^2$ minimization approach. The line peaks at rest-frame $1525\pm4.4\,\ang$ (i.e., it is blueshifted by $\Delta v\approx-4565\pm859\,\mathrm{km\,s^{-1}}$), with a $FWHM=9063\pm2040\,\mathrm{km\,s^{-1}}$ and rest-frame equivalent width (REW) of $32^{+15}_{-12}\,\ang$. Absorption features might be present blueward of the Si IV and C IV emission lines, but a spectrum with higher SNR is required to test this scenario.

\subsubsection{Mg II emission line }

The Mg II ($2798\,\ang$) emission line is an important feature in the spectra of high-redshift QSOs, as it is often used to obtain single-epoch estimates of the SMBH mass \citep[e.g.,][]{Vestergaard09,Shen11}. Based on their analysis of the 2014 spectrum of PSO167--13, \cite{Mazzucchelli17b}  reported a $\mathrm{FWHM_{MgII}}=2071^{+211}_{-354}\,\mathrm{km\,s^{-1}}$, and estimated $M_{BH}=3\times10^8\,\mathrm{M_\odot}$ using the calibration of \cite{Vestergaard09}:

\begin{equation}
M_{BH}=10^{6.86}\left(\frac{FWHM}{10^3\,\mathrm{km\, s^{-1}}}\right)^2\left(\frac{\lambda L_{\lambda 3000}}{10^{44}\,\mathrm{erg s^{-1}}}\right)^{0.5} \mathrm{M_\odot}
\end{equation}

\begin{figure}
	\begin{center}
		\hbox{
			\includegraphics[width=83mm,keepaspectratio]{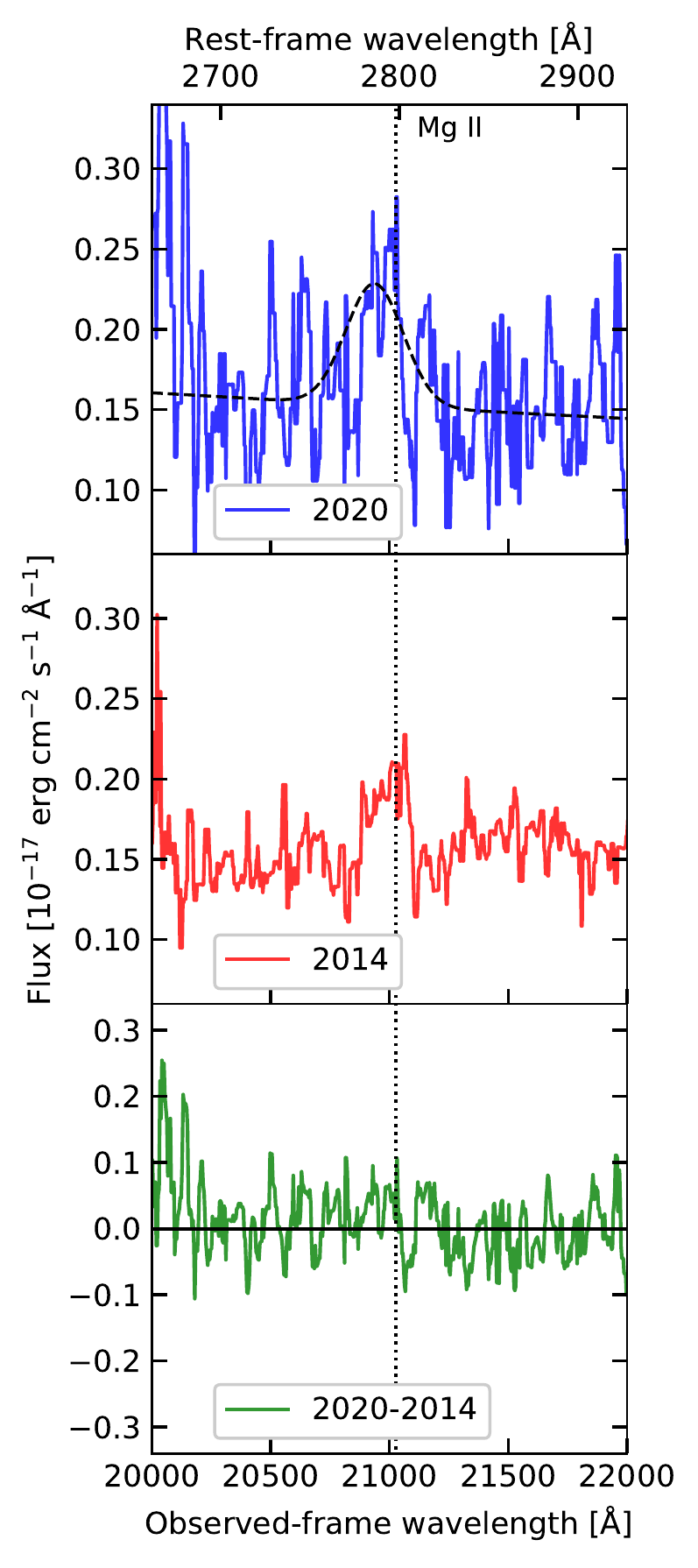} 
		}
	\end{center}
	\caption{ Zoom of the spectral region of the 2020 (top panel) and 2014 (middle panel) spectra encompassing the Mg II emission line of PSO167--13. The bottom panel presents the difference of the two epochs. The dashed black line in the top panel marks the best-fitting continuum plus single Gaussian model of the Mg II emission line.
	}\label{Fig_MgII}
\end{figure}

Fig.~\ref{Fig_MgII} shows the PSO167--13 spectrum in the $2.0-2.2\,\mathrm\um$ spectral region. We fit the Mg II line  assuming a QSO UV power-law continuum (Eq.~\ref{UV_cont}), the Balmer pseudo-continuum modelled as in \cite{Schindler20}, 
the iron pseudo-continuum template of \cite{Vestergaard01}, convolved with a Gaussian function with $\sigma$ equal to that of the best-fitting Mg II line (see, e.g., \citealt{Vestergaard01},  \citealt{Schindler20}),
and a single Gaussian function.  The best-fitting model is shown as a dashed blue line in Fig.~\ref{Fig_MgII}. The Gaussian is centered at rest-frame $\lambda=2786\pm3\,\ang$; i.e., $\Delta v\approx-1268\pm306\,\mathrm{km\,s^{-1}}$ from the expected position given by the [C II] systemic redshift.
The fit returns a $REW_{MgII}=19^{+8}_{-6}\,\ang$, which is consistent with typical values for $z>6$ QSOs \citep[e.g.,][]{Onoue20,Schindler20}.

The Mg II width (FWHM$_{MgII}=3947\pm758\,\mathrm{km\,s^{-1}}$) is significantly larger than the value found by \cite{Mazzucchelli17b}, and translates into $M_{BH}=1.1\times10^9\,\mathrm{M_\odot}$\footnote{Errors on single-epoch black-hole mass estimates are dominated by systematic uncertainties of the calibration ($>0.5$ dex; e.g., \citealt{Shen13} and references therein). } and an Eddington ratio $\lambda_{Edd}=0.3$. We caution that the line's blueshift may indicate the presence of outflowing nuclear winds (see \S~\ref{winds}), in which case the virial assumption upon which  the BH mass estimate is based would be affected. In addition, the limited spectral quality and the different BH mass value obtained from the 2014 spectrum lead us to refrain from over-interpreting these results.

\section{Discussion}

\subsection{X-ray emission from the companion galaxy}
In  \cite{Vito19a}, we discussed the detection of a faint and hard X-ray source in the PSO167--13 system, consistent with the position of the companion galaxy, although the relatively large positional uncertainty prevented us from discarding an association with the optical QSO. If the X-ray source had been confirmed with the new \chandra observations and an unambiguous association with the companion galaxy had been obtained, it would have been considered a (obscured) QSO, and the PSO167--13 system would have been a QSO pair at high redshift.

Considering the total $177$ ks coverage (\S~\ref{galaxy}), the significance of the emission in the full band ($P=0.985$) is slightly below the detection threshold used in \citet[i.e.; $P=0.99$]{Vito19a,Vito19b}, and is dominated by the three counts that were already reported in \cite{Vito19a}. We note that in that work we used a narrower energy band for detection (i.e., 2--5 keV). Repeating the detection procedure in that band including the new \chandra observations, the emission is still nominally significant ($P=0.992$), with $2.6_{-1.4}^{+2.1}$ net counts, implying $F_{\mathrm{2-5\,\mathrm{keV}}}=1.38_{-0.74}^{+1.11}\times10^{-16}\,\mathrm{erg\,cm^{-2}\,s^{-1}}$. 

 Based on the three net counts detected in 59.3 ks in the 2018 pointing \citep{Vito19a}, we would have expected to detect $5.7$ counts in the new observations (117.4 ks), if the detection were real (i.e. not due to spurious background emission)  and not highly variable. Instead, we detected none, corresponding to a Poisson probability $P(k=0,\mu=5.7)=0.003$. Therefore, either the counts detected in the 2018 observation were due to a strong background fluctuation (in \citealt{Vito19a} we assigned a probability of $4\times10^{-4}$ to this possibility), or the \mbox{X-ray} source is variable at $0.997$ significance. In the latter case, comparing the flux derived in \citet[i.e., $F_{2-5\,\mathrm{keV}}=8^{+6}_{-4}\times10^{-16}\,\mathrm{erg\,cm^{-2}\,s^{-1}}$]{Vito19a} with the flux computed from the new 117.4 ks only (i.e., $F_{2-5\,\mathrm{keV}}<2\times10^{-16}\,\mathrm{erg\,cm^{-2}\,s^{-1}}$), the X-ray source would have dimmed by a factor of $\approx4$ from 2018 to 2020; i.e., about three months in the QSO rest frame.
A similar variability (i.e., a factor of $\approx2.5$ in flux over a period of $\approx2$ years in the rest frame) was reported by \cite{Nanni17} for another $z>6$ QSO, namely SDSS J1030+0524 (see also \citealt{Shemmer05} for a  $z=5.41$ SDSS QSO). 

Whatever the cause, we do not confirm the presence of  significant X-ray emission from the PSO167--13 system, and in particular from the companion galaxy. 
To date, out of the $\approx20$ companion galaxies currently detected with \textit{ALMA} close to $z>6$ QSOs \citep{Willott17,Decarli17, Decarli18, Neeleman19,Venemans20}, none has been detected in the X-rays with high significance (i.e., $P>0.99$; see also \citealt{Connor19, Connor20}).

\subsection{X-ray weakness of PSO167--13}\label{Xrayweakness}
The X-ray weakness of PSO167–13 discussed in \S~\ref{PSO167}, i.e., $> 6$ weaker than expected based on its UV luminosity, is notable.
 BAL QSOs have been generally found to be X-ray weak, by factors up to $\approx100$ in extreme cases (e.g., \citealt{Gallagher06, Gibson09a,Luo14,Liu18}), which is possibly linked to the acceleration of nuclear winds in these objects (see \S~\ref{winds}).
 
Among the general SDSS radio-quiet and non-BAL QSO population, \cite{Pu20} found that only $\approx6\%$ show similar levels of X-ray weakness, and they are preferentially WLQs or red QSOs (see also, e.g., \citealt{Ni18, Timlin20}). 
Applying this fraction to the total number of $\bm{z>6}$ QSOs currently observed in the X-ray band  (i.e., 36; \citealt{Vito19b} and references therein, \citealt{Connor19,Connor20,Pons20,Wang21a}) we expect $\approx2$ of them to be X-ray weak by  factors greater than six. However, we note that some $z>6$ QSOs with X-ray observations are BAL QSOs (e.g., \citealt{Fan03, Matsuoka16, Connor20}). Hence, they should not be considered here for a proper comparison with the \cite{Pu20} sample,  which includes only radio-quiet and non-BAL QSOs, thus decreasing the expected number of X-ray weak QSOs at $z>6$. Given the quality of the currently available rest-frame spectrum, PSO167--13 cannot be securely identified either as a WLQ or BAL QSO, although there might be evidence for the presence of nuclear winds (see \S~\ref{winds}).

Several physical reasons can be invoked to explain the lack of strong X-ray emission from a type 1 QSO:

 \begin{enumerate}
	\item Intrinsic X-ray weakness may be caused by a different geometry or physics of the hot corona from those of typical QSOs. For instance, the hot corona may be quenched, disrupted, or not yet formed, due to currently not well understood causes (e.g., \citealt{Leighly07}, \citealt{Luo13}, \citealt{Luo14}, \citealt{Liu18}). In this case, the accretion disk can remain largely unaffected, and emit UV photons as a typical QSO. 
	
\item 	Absorption on scales smaller than most of the accretion disk (e.g., due to a thick inner disk, a ``failed wind", or the outflowing material itself in the case of radiation-pressure confinement;  e.g., \citealt{Proga04,Baskin14,Ni18,Liu21}) can absorb the X-ray photons from the hot corona, leaving the UV emission from the accretion disk largely unaffected, thus resulting in significant X-ray weakness observed for type 1 QSOs.

\item  Intrinsic variability can cause a QSO to be observed during periods of low X-ray flux state \citep[e.g.,][and references therein]{Pu20}, although luminous QSOs often do not show large variability amplitudes \citep[e.g.,][]{Paolillo17, Shemmer17}. For instance, \cite{Timlin20b} found that QSOs with UV luminosity similar to that of PSO167--13 vary by a factor of $<3$.

\item Occultation events of broad emission-line clouds with angular size comparable to that of the hot corona can obscure the X-ray emission, although they are usually found to happen on shorter rest-frame timescales (a few hours) than those we probe for PSO167--13 \citep[e.g.,][]{Risaliti11, DeMarco20}.
	
\end{enumerate}

Assuming that the X-ray emission of PSO167--13 is absorbed by intervening neutral material with solar metallicity, and that the intrinsic X-ray luminosity is consistent with the $\alpha_{ox}-L_{UV}$ relation (i.e., $L_{2-10\,\mathrm{keV}}=5.1\times10^{44}\,\mathrm{erg\,s^{-1}}$), we used the self-consistent MYTorus model \citep{Murphy09} in XSPEC to estimate the equivalent hydrogen column density required to match the upper limits on flux reported in \S~\ref{PSO167}. We fixed the intrinsic powerlaw slope to $\Gamma=2$, the normalization of the scattered and line components to that of the transmitted component, and the inclination angle to 90 deg. We found that absorption due to Compton-thick ($N_H\gtrsim10^{24}\,\mathrm{cm^{-2}}$) material is required. However, the physical and geometrical assumptions of the MYTorus model might not be a good representation of the obscuring material in the inner regions on PSO167--13, which is expected to lie on smaller scales than the accretion disk to allow for the detection of the UV emission. Using a simple absorbed powerlaw model in XSPEC (model \textit{zwabs} $\times$ \textit{powerlaw}),  which however does not include a treatment of photon scattering, the column densities required to match the observed fluxes in the soft and hard bands are $N_H\gtrsim2\times10^{24}\,\mathrm{cm^{-2}}$ and $N_H\gtrsim9\times10^{24}\,\mathrm{cm^{-2}}$, respectively. Constraining such high values of $N_H$ is possible thanks to the high-redshift nature of this QSO, which shifts the photoelectric cut-off to low observed energies even for large column densities,\footnote{The \chandra bandpass samples the rest-frame energy range \mbox{4--50~keV} at $z=6.515$. } its relatively high UV luminosity (and, hence, expected X-ray luminosity), and the depth of the available \chandra observations. We conclude that the most plausible causes of the lack of strong \mbox{X-ray} emission from PSO167--13 are either intrinsic X-ray  weakness, possibly due to an accretion mechanism different from that of typical QSOs,  or small-scale absorption by Compton thick material.

\subsection{Possible nuclear winds in PSO167--13}\label{winds}

Both the C IV and Mg II emission lines in the UV spectrum of PSO167--13 show large tentative blueshifts with respect to the [C II] systemic redshift ($-4565\pm859\,\mathrm{km\,s^{-1}}$ and $-1268\pm306\,\mathrm{km\,s^{-1}}$, respectively), although the derivation of accurate physical parameters for these lines is affected by the low SNR of the spectrum, and, in particular for the C IV line, by the low atmospheric transmission. We also note that absorption features blueward of the C IV line may be present, similar to the blueshifted features in BAL QSOs.
Rapidly accreting QSOs, WLQs (which are thought to be accreting close to the Eddington limit), and, in particular, high-redshift QSOs often exhibit similarly large or even larger C IV and Mg II blueshifts \citep[ e.g.,][Yi et al. in prep]{Luo15, Plotkin15, Vietri18, Vietri20, Venemans16,Ni18, Yi19, Onoue20, Schindler20}, which are usually considered to be produced by outflowing winds.

In this respect, the X-ray weakness of PSO167--13 (either intrinsic or due to small-scale obscuration)
may play an important role in the acceleration mechanisms of such winds. In fact, \mbox{X-ray} weakness can help avoid the overionization of the accreting gas, thus allowing efficient launching of UV-line driven winds \citep[e.g., ][]{Proga00,Proga04,Baskin14}. Therefore, it is perhaps not surprising that a relation has been found between $\alpha_{ox}$ and the blueshift of the C IV emission line \citep[e.g.,][]{Richards11,Timlin20,Vietri20}, in the sense that objects with larger C IV blueshifts have softer (i.e., UV dominated) spectra. For instance, the values of $\alpha_{ox}$ and C IV blueshift of PSO167--13 are consistent with the relations found by \cite{Timlin20} and \cite{Zappacosta20}, although for QSOs with $\approx1$ dex higher bolometric luminosities.

While the C IV emission line in QSOs is usually found to be blueshifted by $\approx1000\,\mathrm{km\,s^{-1}}$ with respect to the Mg II line \citep[e.g.,][]{Meyer19}, evidence has been found recently  for an increasing $\Delta v(\mathrm{CIV - MgII})$ at $z>6$ \citep[][but see also \citealt{Shen19}]{Meyer19,Schindler20}, with an average value of $\approx-3000\,\mathrm{km\,s^{-1}}$ and up $\approx-5000\,\mathrm{km\,s^{-1}}$ at $z=6.5$. For PSO167--13, we found $\Delta v(\mathrm{CIV - MgII})=-3300\,\mathrm{km\,s^{-1}}$, in agreement with the results of \cite{Schindler20} at similar redshift, but for QSOs which are typically $\approx1$ dex more luminous in the UV.
A higher signal-to-noise NIR spectrum is required to confirm the tentative nature of the rest-frame UV line properties of this PSO167–-13.
 
 \section{Conclusions}
We present deep X-ray (\chandra, 177 ks in total) and NIR spectroscopic (\textit{Magellan}/FIRE, 7.2h on source) follow-up observations of PSO167--13, an optically selected $z=6.515$ QSO ($M_{1450\ang}=-25.6$) in an interacting system with a close ($0.9^{\prime\prime}$, corresponding to $\approx5$ projected kpc)  companion galaxy detected with both ALMA and HST. A previous tentative detection  of a hard X-ray source with \chandra ($59.3$ ks) suggested the presence of obscured nuclear accretion in this system. We summarize here the main results:

\begin{itemize}
	\item The new \chandra observations do not confirm significant \mbox{X-ray} emission from the QSO-galaxy system, suggesting that the previously detected X-ray source was due to a strong background fluctuation, although intrinsic variability by a factor $\approx4$ cannot be excluded.
	
		\item We calculate upper limits (at the 90\% confidence level) on the X-ray flux of the companion galaxy ($F<(1.9/6.6/5.0)\times10^{-16}\,\mathrm{erg\,cm^{-2}\,s^{-1}}$ in the soft/hard/full bands) and the intrinsic 2--10 keV luminosity ($L_{X}<1.3\times10^{44}\,\mathrm{erg\,s^{-1}}$). To date, none of the ALMA detected companion galaxies in the proximity of $z>6$ QSOs has been detected with high significance ($P>0.99$) in standard X-ray bands.
	
	\item Likewise, we place upper limits on the X-ray flux from PSO167--13 of $F<(1.2/5.1/3.2)\times10^{-16}\,\mathrm{erg\,cm^{-2}\,s^{-1}}$ in the soft/hard/full bands and intrinsic luminosity \mbox{$L_{2-10\,\mathrm{keV}}<8.3\times10^{43}\,\mathrm{erg\,s^{-1}}$}. These are the lowest upper limits on the X-ray emission for a  $z>6$ QSO.
	
	\item The ratio between the X-ray and UV luminosity  of PSO167--13, $\alpha_{ox}<-1.95$, makes PSO167--13 an outlier from the \mbox{$\alpha_{ox}-L_{UV} $} relation for QSOs, with a deviation of \mbox{$\Delta\alpha_{ox}<-0.30$}, corresponding to a factor $>6$ weaker X-ray emission than the expectation. Only $\approx6\%$ of SDSS radio-quiet non-BAL QSOs show similar X-ray weakness, and they are usually WLQs or red QSOs. 
	Such weak X-ray emission for PSO167--13 could be intrinsic (e.g., due to an accretion configuration different from typical optically selected QSOs), or due to small-scale obscuration, which would allow the detection of the UV continuum. In the latter case, we estimate a column density of $N_H>10^{24}\,\mathrm{cm^{-2}}$.
	
	\item The slope of the rest-frame UV spectrum of PSO167--13 taken in 2020 and presented here ($\alpha=-1.10\pm0.12$) is consistent with previous spectroscopy, and redder than typical values for optically selected QSOs. Absorption features may be present blueward of the C IV line, but the low SNR of the spectrum prevents their definitive assessment. 
	
	\item The tentatively detected C IV and Mg II emission lines appear to be broad ($FWHM=9063\pm2040\,\mathrm{km\,s^{-1}}$ and $3947\pm758\,\mathrm{km\,s^{-1}}$, respectively) and strongly blueshifted from the systemic redshift based on the [C II] 158$\mu m$ line ($\Delta v=-4565\,\mathrm{km\,s^{-1}}$ and $-1268\,\mathrm{km\,s^{-1}}$, respectively). Similar large blueshifts have been found in other $z>6$ QSOs, and in rapidly accreting QSOs and WLQs at lower redshifts, and are generally associated with the presence of nuclear winds. The C IV line is found to be blueshifted with respect to the Mg II line by $\Delta v(CIV-MgII)=-3300\,\mathrm{km\,s^{-1}}$. This value is consistent with recent findings for $z>6$ QSOs. However, we note that the spectroscopic observations were taken during nights with poor seeing and strongly varying atmospheric conditions. A higher signal-to-noise NIR spectrum is required to confirm the tentative nature of the rest-frame UV line properties of this PSO167–-13.
	
	\item As suggested by the relation between $\alpha_{ox}$ and C IV blueshift found by previous works, the unusual X-ray weakness of PSO167--13 might facilitate the acceleration of such winds by preventing the overionization of the accreting material, which is required by models of UV-driven wind acceleration. Based on the $FWHM$ of the Mg II line, we estimate a virial BH mass of $1.1\times10^{9}\,M_\odot$, corresponding to $\lambda_{Edd}=0.3$, but we caution that the presence of nuclear winds could severely affect this measurement.
\end{itemize}

 \begin{acknowledgements}
 We thank the referee, Belinda Wilkes, for her useful comments and suggestions.
We thank Marcel Neeleman, Bram Venemans, Estelle Pons, and Weimin Yi for useful discussions, and Marianne Vestergaard for providing the iron UV emission template of \cite{Vestergaard01}. 
We acknowledge support from CXC grants GO0-21078D (W.N.B.) and GO0-21078C (O.S.), from ANID grants CATA-Basal AFB-170002 (F.E.B., E.C.), 
FONDECYT Regular 1190818 (F.E.B.) and 1200495 (F.E.B.),
Millennium Science Initiative ICN12\_009 (F.E.B.), the NSFC grant 11991053 and National Key R\&D Program of China grant 2016YFA0400702 (B.L.), from ASI-INAF n. 2018-31-HH.0 grant and PRIN-MIUR 2017 (S.G.), and from the agreement ASI-INAF n. 2017-14-H.O.
The work of T.C. was carried out at the Jet Propulsion Laboratory, California Institute of Technology, under a contract with NASA. 
This research has made use of data obtained from the Chandra Data Archive (Proposal IDs 19700183 and 21700027), and software provided by the Chandra X-ray Center (CXC) in the application packages CIAO. This paper includes data gathered with the 6.5 meter Magellan Telescopes located at Las Campanas Observatory, Chile (CNTAC proposal ID CN2020A-22). This research made use of SAO Image DS9 \cite{Joye03} and Astropy,\footnote{\url{http://www.astropy.org}} a community-developed core Python package for Astronomy \citep{Astropy13,Astropy18}.

\end{acknowledgements}

\bibliographystyle{aa}
\bibliography{biblio.bib} % if your bibtex file is called example.bib

\begin{thebibliography}{111}
\expandafter\ifx\csname natexlab\endcsname\relax\def\natexlab#1{#1}\fi

\bibitem[{{Astropy Collaboration} {et~al.}(2018){Astropy Collaboration},
  {Price-Whelan}, {Sip{\H{o}}cz}, {G{\"u}nther}, {Lim}, {Crawford}, {Conseil},
  {Shupe}, {Craig}, {Dencheva}, {Ginsburg}, {Vand erPlas}, {Bradley},
  {P{\'e}rez-Su{\'a}rez}, {de Val-Borro}, {Aldcroft}, {Cruz}, {Robitaille},
  {Tollerud}, {Ardelean}, {Babej}, {Bach}, {Bachetti}, {Bakanov}, {Bamford},
  {Barentsen}, {Barmby}, {Baumbach}, {Berry}, {Biscani}, {Boquien}, {Bostroem},
  {Bouma}, {Brammer}, {Bray}, {Breytenbach}, {Buddelmeijer}, {Burke},
  {Calderone}, {Cano Rodr{\'\i}guez}, {Cara}, {Cardoso}, {Cheedella}, {Copin},
  {Corrales}, {Crichton}, {D'Avella}, {Deil}, {Depagne}, {Dietrich}, {Donath},
  {Droettboom}, {Earl}, {Erben}, {Fabbro}, {Ferreira}, {Finethy}, {Fox},
  {Garrison}, {Gibbons}, {Goldstein}, {Gommers}, {Greco}, {Greenfield},
  {Groener}, {Grollier}, {Hagen}, {Hirst}, {Homeier}, {Horton}, {Hosseinzadeh},
  {Hu}, {Hunkeler}, {Ivezi{\'c}}, {Jain}, {Jenness}, {Kanarek}, {Kendrew},
  {Kern}, {Kerzendorf}, {Khvalko}, {King}, {Kirkby}, {Kulkarni}, {Kumar},
  {Lee}, {Lenz}, {Littlefair}, {Ma}, {Macleod}, {Mastropietro}, {McCully},
  {Montagnac}, {Morris}, {Mueller}, {Mumford}, {Muna}, {Murphy}, {Nelson},
  {Nguyen}, {Ninan}, {N{\"o}the}, {Ogaz}, {Oh}, {Parejko}, {Parley}, {Pascual},
  {Patil}, {Patil}, {Plunkett}, {Prochaska}, {Rastogi}, {Reddy Janga},
  {Sabater}, {Sakurikar}, {Seifert}, {Sherbert}, {Sherwood-Taylor}, {Shih},
  {Sick}, {Silbiger}, {Singanamalla}, {Singer}, {Sladen}, {Sooley},
  {Sornarajah}, {Streicher}, {Teuben}, {Thomas}, {Tremblay}, {Turner},
  {Terr{\'o}n}, {van Kerkwijk}, {de la Vega}, {Watkins}, {Weaver}, {Whitmore},
  {Woillez}, {Zabalza}, \& {Astropy Contributors}}]{Astropy18}
{Astropy Collaboration}, {Price-Whelan}, A.~M., {Sip{\H{o}}cz}, B.~M., {et~al.}
  2018, \aj, 156, 123

\bibitem[{{Astropy Collaboration} {et~al.}(2013){Astropy Collaboration},
  {Robitaille}, {Tollerud}, {Greenfield}, {Droettboom}, {Bray}, {Aldcroft},
  {Davis}, {Ginsburg}, {Price-Whelan}, {Kerzendorf}, {Conley}, {Crighton},
  {Barbary}, {Muna}, {Ferguson}, {Grollier}, {Parikh}, {Nair}, {Unther},
  {Deil}, {Woillez}, {Conseil}, {Kramer}, {Turner}, {Singer}, {Fox}, {Weaver},
  {Zabalza}, {Edwards}, {Azalee Bostroem}, {Burke}, {Casey}, {Crawford},
  {Dencheva}, {Ely}, {Jenness}, {Labrie}, {Lim}, {Pierfederici}, {Pontzen},
  {Ptak}, {Refsdal}, {Servillat}, \& {Streicher}}]{Astropy13}
{Astropy Collaboration}, {Robitaille}, T.~P., {Tollerud}, E.~J., {et~al.} 2013,
  \aap, 558, A33

\bibitem[{{Ba{\~n}ados} {et~al.}(2018{\natexlab{a}}){Ba{\~n}ados}, {Connor},
  {Stern}, {Mulchaey}, {Fan}, {Decarli}, {Farina}, {Mazzucchelli}, {Venemans},
  {Walter}, {Wang}, \& {Yang}}]{Banados18b}
{Ba{\~n}ados}, E., {Connor}, T., {Stern}, D., {et~al.} 2018{\natexlab{a}},
  \apjl, 856, L25

\bibitem[{{Ba{\~n}ados} {et~al.}(2016){Ba{\~n}ados}, {Venemans}, {Decarli},
  {Farina}, {Mazzucchelli}, {Walter}, {Fan}, {Stern}, {Schlafly}, {Chambers},
  {Rix}, {Jiang}, {McGreer}, {Simcoe}, {Wang}, {Yang}, {Morganson}, {De Rosa},
  {Greiner}, {Balokovi{\'c}}, {Burgett}, {Cooper}, {Draper}, {Flewelling},
  {Hodapp}, {Jun}, {Kaiser}, {Kudritzki}, {Magnier}, {Metcalfe}, {Miller},
  {Schindler}, {Tonry}, {Wainscoat}, {Waters}, \& {Yang}}]{Banados16}
{Ba{\~n}ados}, E., {Venemans}, B.~P., {Decarli}, R., {et~al.} 2016, \apjs, 227,
  11

\bibitem[{{Ba{\~n}ados} {et~al.}(2018{\natexlab{b}}){Ba{\~n}ados}, {Venemans},
  {Mazzucchelli}, {Farina}, {Walter}, {Wang}, {Decarli}, {Stern}, {Fan},
  {Davies}, {Hennawi}, {Simcoe}, {Turner}, {Rix}, {Yang}, {Kelson}, {Rudie}, \&
  {Winters}}]{Banados18a}
{Ba{\~n}ados}, E., {Venemans}, B.~P., {Mazzucchelli}, C., {et~al.}
  2018{\natexlab{b}}, \nat, 553, 473

\bibitem[{{Baskin} {et~al.}(2014){Baskin}, {Laor}, \& {Stern}}]{Baskin14}
{Baskin}, A., {Laor}, A., \& {Stern}, J. 2014, \mnras, 445, 3025

\bibitem[{{Belladitta} {et~al.}(2020){Belladitta}, {Moretti}, {Caccianiga},
  {Spingola}, {Severgnini}, {Della Ceca}, {Ghisellini}, {Dallacasa},
  {Sbarrato}, {Cicone}, {Cassar{\`a}}, \& {Pedani}}]{Belladitta20}
{Belladitta}, S., {Moretti}, A., {Caccianiga}, A., {et~al.} 2020, \aap, 635, L7

\bibitem[{{Brandt} \& {Alexander}(2015)}]{Brandt15}
{Brandt}, W.~N. \& {Alexander}, D.~M. 2015, \aapr, 23, 1

\bibitem[{{Brightman} {et~al.}(2013){Brightman}, {Silverman}, {Mainieri},
  {Ueda}, {Schramm}, {Matsuoka}, {Nagao}, {Steinhardt}, {Kartaltepe},
  {Sanders}, {Treister}, {Shemmer}, {Brandt}, {Brusa}, {Comastri}, {Ho},
  {Lanzuisi}, {Lusso}, {Nandra}, {Salvato}, {Zamorani}, {Akiyama}, {Alexander},
  {Bongiorno}, {Capak}, {Civano}, {Del Moro}, {Doi}, {Elvis}, {Hasinger},
  {Laird}, {Masters}, {Mignoli}, {Ohta}, {Schawinski}, \&
  {Taniguchi}}]{Brightman13}
{Brightman}, M., {Silverman}, J.~D., {Mainieri}, V., {et~al.} 2013, \mnras,
  433, 2485

\bibitem[{{Broos} {et~al.}(2007){Broos}, {Feigelson}, {Townsley}, {Getman},
  {Wang}, {Garmire}, {Jiang}, \& {Tsuboi}}]{Broos07}
{Broos}, P.~S., {Feigelson}, E.~D., {Townsley}, L.~K., {et~al.} 2007, \apjs,
  169, 353

\bibitem[{{Chambers} {et~al.}(2016){Chambers}, {Magnier}, {Metcalfe},
  {Flewelling}, {Huber}, {Waters}, {Denneau}, {Draper}, {Farrow}, {Finkbeiner},
  {Holmberg}, {Koppenhoefer}, {Price}, {Rest}, {Saglia}, {Schlafly}, {Smartt},
  {Sweeney}, {Wainscoat}, {Burgett}, {Chastel}, {Grav}, {Heasley}, {Hodapp},
  {Jedicke}, {Kaiser}, {Kudritzki}, {Luppino}, {Lupton}, {Monet}, {Morgan},
  {Onaka}, {Shiao}, {Stubbs}, {Tonry}, {White}, {Ba{\~n}ados}, {Bell},
  {Bender}, {Bernard}, {Boegner}, {Boffi}, {Botticella}, {Calamida},
  {Casertano}, {Chen}, {Chen}, {Cole}, {Deacon}, {Frenk}, {Fitzsimmons},
  {Gezari}, {Gibbs}, {Goessl}, {Goggia}, {Gourgue}, {Goldman}, {Grant},
  {Grebel}, {Hambly}, {Hasinger}, {Heavens}, {Heckman}, {Henderson}, {Henning},
  {Holman}, {Hopp}, {Ip}, {Isani}, {Jackson}, {Keyes}, {Koekemoer}, {Kotak},
  {Le}, {Liska}, {Long}, {Lucey}, {Liu}, {Martin}, {Masci}, {McLean}, {Mindel},
  {Misra}, {Morganson}, {Murphy}, {Obaika}, {Narayan}, {Nieto-Santisteban},
  {Norberg}, {Peacock}, {Pier}, {Postman}, {Primak}, {Rae}, {Rai}, {Riess},
  {Riffeser}, {Rix}, {R{\"o}ser}, {Russel}, {Rutz}, {Schilbach}, {Schultz},
  {Scolnic}, {Strolger}, {Szalay}, {Seitz}, {Small}, {Smith}, {Soderblom},
  {Taylor}, {Thomson}, {Taylor}, {Thakar}, {Thiel}, {Thilker}, {Unger},
  {Urata}, {Valenti}, {Wagner}, {Walder}, {Walter}, {Watters}, {Werner},
  {Wood-Vasey}, \& {Wyse}}]{Chambers16}
{Chambers}, K.~C., {Magnier}, E.~A., {Metcalfe}, N., {et~al.} 2016, arXiv
  e-prints, arXiv:1612.05560

\bibitem[{{Connor} {et~al.}(2020){Connor}, {Ba{\~n}ados}, {Mazzucchelli},
  {Stern}, {Decarli}, {Fan}, {Farina}, {Lusso}, {Neeleman}, \&
  {Walter}}]{Connor20}
{Connor}, T., {Ba{\~n}ados}, E., {Mazzucchelli}, C., {et~al.} 2020, \apj, 900,
  189

\bibitem[{{Connor} {et~al.}(2019){Connor}, {Ba{\~n}ados}, {Stern}, {Decarli},
  {Schindler}, {Fan}, {Farina}, {Mazzucchelli}, {Mulchaey}, \&
  {Walter}}]{Connor19}
{Connor}, T., {Ba{\~n}ados}, E., {Stern}, D., {et~al.} 2019, \apj, 887, 171

\bibitem[{{Davies} {et~al.}(2019){Davies}, {Hennawi}, \& {Eilers}}]{Davies19}
{Davies}, F.~B., {Hennawi}, J.~F., \& {Eilers}, A.-C. 2019, \apjl, 884, L19

\bibitem[{{De Marco} {et~al.}(2020){De Marco}, {Adhikari}, {Ponti}, {Bianchi},
  {Kriss}, {Arav}, {Behar}, {Branduardi-Raymont}, {Cappi}, {Costantini},
  {Costanzo}, {di Gesu}, {Ebrero}, {Kaastra}, {Kaspi}, {Mao}, {Markowitz},
  {Matt}, {Mehdipour}, {Middei}, {Paltani}, {Petrucci}, {Pinto},
  {R{\'o}{\.z}a{\'n}ska}, \& {Walton}}]{DeMarco20}
{De Marco}, B., {Adhikari}, T.~P., {Ponti}, G., {et~al.} 2020, \aap, 634, A65

\bibitem[{{De Rosa} {et~al.}(2014){De Rosa}, {Venemans}, {Decarli}, {Gennaro},
  {Simcoe}, {Dietrich}, {Peterson}, {Walter}, {Frank}, {McMahon}, {Hewett},
  {Mortlock}, \& {Simpson}}]{DeRosa14}
{De Rosa}, G., {Venemans}, B.~P., {Decarli}, R., {et~al.} 2014, \apj, 790, 145

\bibitem[{{Decarli} {et~al.}(2019){Decarli}, {Dotti}, {Ba{\~n}ados}, {Farina},
  {Walter}, {Carilli}, {Fan}, {Mazzucchelli}, {Neeleman}, {Novak}, {Riechers},
  {Strauss}, {Venemans}, {Yang}, \& {Wang}}]{Decarli19}
{Decarli}, R., {Dotti}, M., {Ba{\~n}ados}, E., {et~al.} 2019, \apj, 880, 157

\bibitem[{{Decarli} {et~al.}(2017){Decarli}, {Walter}, {Venemans},
  {Ba{\~n}ados}, {Bertoldi}, {Carilli}, {Fan}, {Farina}, {Mazzucchelli},
  {Riechers}, {Rix}, {Strauss}, {Wang}, \& {Yang}}]{Decarli17}
{Decarli}, R., {Walter}, F., {Venemans}, B.~P., {et~al.} 2017, \nat, 545, 457

\bibitem[{{Decarli} {et~al.}(2018){Decarli}, {Walter}, {Venemans},
  {Ba{\~n}ados}, {Bertoldi}, {Carilli}, {Fan}, {Farina}, {Mazzucchelli},
  {Riechers}, {Rix}, {Strauss}, {Wang}, \& {Yang}}]{Decarli18}
{Decarli}, R., {Walter}, F., {Venemans}, B.~P., {et~al.} 2018, \apj, 854, 97

\bibitem[{{Duras} {et~al.}(2020){Duras}, {Bongiorno}, {Ricci}, {Piconcelli},
  {Shankar}, {Lusso}, {Bianchi}, {Fiore}, {Maiolino}, {Marconi}, {Onori},
  {Sani}, {Schneider}, {Vignali}, \& {La Franca}}]{Duras20}
{Duras}, F., {Bongiorno}, A., {Ricci}, F., {et~al.} 2020, \aap, 636, A73

\bibitem[{{Eilers} {et~al.}(2018){Eilers}, {Hennawi}, \& {Davies}}]{Eilers18}
{Eilers}, A.-C., {Hennawi}, J.~F., \& {Davies}, F.~B. 2018, \apj, 867, 30

\bibitem[{{Eilers} {et~al.}(2020){Eilers}, {Hennawi}, {Decarli}, {Davies},
  {Venemans}, {Walter}, {Ba{\~n}ados}, {Fan}, {Farina}, {Mazzucchelli},
  {Novak}, {Schindler}, {Simcoe}, {Wang}, \& {Yang}}]{Eilers20}
{Eilers}, A.-C., {Hennawi}, J.~F., {Decarli}, R., {et~al.} 2020, \apj, 900, 37

\bibitem[{{Fan} {et~al.}(2003){Fan}, {Strauss}, {Schneider}, {Becker}, {White},
  {Haiman}, {Gregg}, {Pentericci}, {Grebel}, {Narayanan}, {Loh}, {Richards},
  {Gunn}, {Lupton}, {Knapp}, {Ivezi{\'c}}, {Brandt}, {Collinge}, {Hao},
  {Harbeck}, {Prada}, {Schaye}, {Strateva}, {Zakamska}, {Anderson},
  {Brinkmann}, {Bahcall}, {Lamb}, {Okamura}, {Szalay}, \& {York}}]{Fan03}
{Fan}, X., {Strauss}, M.~A., {Schneider}, D.~P., {et~al.} 2003, \aj, 125, 1649

\bibitem[{{Fan} {et~al.}(2019){Fan}, {Wang}, {Yang}, {Keeton}, {Yue},
  {Zabludoff}, {Bian}, {Bonaglia}, {Georgiev}, {Hennawi}, {Li}, {McGreer},
  {Naidu}, {Pacucci}, {Rabien}, {Thompson}, {Venemans}, {Walter}, {Wang}, \&
  {Wu}}]{Fan19}
{Fan}, X., {Wang}, F., {Yang}, J., {et~al.} 2019, \apj, 870, L11

\bibitem[{{Farina} {et~al.}(2019){Farina}, {Arrigoni-Battaia}, {Costa},
  {Walter}, {Hennawi}, {Drake}, {Decarli}, {Gutcke}, {Mazzucchelli},
  {Neeleman}, {Georgiev}, {Eilers}, {Davies}, {Ba{\~n}ados}, {Fan}, {Onoue},
  {Schindler}, {Venemans}, {Wang}, {Yang}, {Rabien}, \& {Busoni}}]{Farina19}
{Farina}, E.~P., {Arrigoni-Battaia}, F., {Costa}, T., {et~al.} 2019, \apj, 887,
  196

\bibitem[{{Fitzpatrick}(1999)}]{Fitzpatrick99}
{Fitzpatrick}, E.~L. 1999, \pasp, 111, 63

\bibitem[{{Fruscione} {et~al.}(2006){Fruscione}, {McDowell}, {Allen},
  {Brickhouse}, {Burke}, {Davis}, {Durham}, {Elvis}, {Galle}, {Harris},
  {Huenemoerder}, {Houck}, {Ishibashi}, {Karovska}, {Nicastro}, {Noble},
  {Nowak}, {Primini}, {Siemiginowska}, {Smith}, \& {Wise}}]{Fruscione06}
{Fruscione}, A., {McDowell}, J.~C., {Allen}, G.~E., {et~al.} 2006, Society of
  Photo-Optical Instrumentation Engineers (SPIE) Conference Series, Vol. 6270,
  {CIAO: Chandra's data analysis system}, 62701V

\bibitem[{{Gagn{\'e}} {et~al.}(2015){Gagn{\'e}}, {Lambrides}, {Faherty}, \&
  {Simcoe}}]{Gagne15}
{Gagn{\'e}}, J., {Lambrides}, E., {Faherty}, J.~K., \& {Simcoe}, R. 2015,
  {Firehose\_V2: Firehose V2.0}

\bibitem[{{Gaia Collaboration} {et~al.}(2018){Gaia Collaboration}, {Brown},
  {Vallenari}, {Prusti}, {de Bruijne}, {Babusiaux}, {Bailer-Jones}, {Biermann},
  {Evans}, {Eyer}, {Jansen}, {Jordi}, {Klioner}, {Lammers}, {Lindegren},
  {Luri}, {Mignard}, {Panem}, {Pourbaix}, {Randich}, {Sartoretti}, {Siddiqui},
  {Soubiran}, {van Leeuwen}, {Walton}, {Arenou}, {Bastian}, {Cropper},
  {Drimmel}, {Katz}, {Lattanzi}, {Bakker}, {Cacciari}, {Casta{\~n}eda},
  {Chaoul}, {Cheek}, {De Angeli}, {Fabricius}, {Guerra}, {Holl}, {Masana},
  {Messineo}, {Mowlavi}, {Nienartowicz}, {Panuzzo}, {Portell}, {Riello},
  {Seabroke}, {Tanga}, {Th{\'e}venin}, {Gracia-Abril}, {Comoretto},
  {Garcia-Reinaldos}, {Teyssier}, {Altmann}, {Andrae}, {Audard},
  {Bellas-Velidis}, {Benson}, {Berthier}, {Blomme}, {Burgess}, {Busso},
  {Carry}, {Cellino}, {Clementini}, {Clotet}, {Creevey}, {Davidson}, {De
  Ridder}, {Delchambre}, {Dell'Oro}, {Ducourant},
  {Fern{\'a}ndez-Hern{\'a}ndez}, {Fouesneau}, {Fr{\'e}mat}, {Galluccio},
  {Garc{\'\i}a-Torres}, {Gonz{\'a}lez-N{\'u}{\~n}ez}, {Gonz{\'a}lez-Vidal},
  {Gosset}, {Guy}, {Halbwachs}, {Hambly}, {Harrison}, {Hern{\'a}ndez},
  {Hestroffer}, {Hodgkin}, {Hutton}, {Jasniewicz}, {Jean-Antoine-Piccolo},
  {Jordan}, {Korn}, {Krone-Martins}, {Lanzafame}, {Lebzelter}, {L{\"o}ffler},
  {Manteiga}, {Marrese}, {Mart{\'\i}n-Fleitas}, {Moitinho}, {Mora}, {Muinonen},
  {Osinde}, {Pancino}, {Pauwels}, {Petit}, {Recio-Blanco}, {Richards},
  {Rimoldini}, {Robin}, {Sarro}, {Siopis}, {Smith}, {Sozzetti}, {S{\"u}veges},
  {Torra}, {van Reeven}, {Abbas}, {Abreu Aramburu}, {Accart}, {Aerts},
  {Altavilla}, {{\'A}lvarez}, {Alvarez}, {Alves}, {Anderson}, {Andrei},
  {Anglada Varela}, {Antiche}, {Antoja}, {Arcay}, {Astraatmadja}, {Bach},
  {Baker}, {Balaguer-N{\'u}{\~n}ez}, {Balm}, {Barache}, {Barata}, {Barbato},
  {Barblan}, {Barklem}, {Barrado}, {Barros}, {Barstow}, {Bartholom{\'e}
  Mu{\~n}oz}, {Bassilana}, {Becciani}, {Bellazzini}, {Berihuete}, {Bertone},
  {Bianchi}, {Bienaym{\'e}}, {Blanco-Cuaresma}, {Boch}, {Boeche}, {Bombrun},
  {Borrachero}, {Bossini}, {Bouquillon}, {Bourda}, {Bragaglia}, {Bramante},
  {Breddels}, {Bressan}, {Brouillet}, {Br{\"u}semeister}, {Brugaletta},
  {Bucciarelli}, {Burlacu}, {Busonero}, {Butkevich}, {Buzzi}, {Caffau},
  {Cancelliere}, {Cannizzaro}, {Cantat-Gaudin}, {Carballo}, {Carlucci},
  {Carrasco}, {Casamiquela}, {Castellani}, {Castro-Ginard}, {Charlot},
  {Chemin}, {Chiavassa}, {Cocozza}, {Costigan}, {Cowell}, {Crifo}, {Crosta},
  {Crowley}, {Cuypers}, {Dafonte}, {Damerdji}, {Dapergolas}, {David}, {David},
  {de Laverny}, {De Luise}, {De March}, {de Martino}, {de Souza}, {de Torres},
  {Debosscher}, {del Pozo}, {Delbo}, {Delgado}, {Delgado}, {Di Matteo},
  {Diakite}, {Diener}, {Distefano}, {Dolding}, {Drazinos}, {Dur{\'a}n},
  {Edvardsson}, {Enke}, {Eriksson}, {Esquej}, {Eynard Bontemps}, {Fabre},
  {Fabrizio}, {Faigler}, {Falc{\~a}o}, {Farr{\`a}s Casas}, {Federici},
  {Fedorets}, {Fernique}, {Figueras}, {Filippi}, {Findeisen}, {Fonti},
  {Fraile}, {Fraser}, {Fr{\'e}zouls}, {Gai}, {Galleti}, {Garabato},
  {Garc{\'\i}a-Sedano}, {Garofalo}, {Garralda}, {Gavel}, {Gavras}, {Gerssen},
  {Geyer}, {Giacobbe}, {Gilmore}, {Girona}, {Giuffrida}, {Glass}, {Gomes},
  {Granvik}, {Gueguen}, {Guerrier}, {Guiraud}, {Guti{\'e}rrez-S{\'a}nchez},
  {Haigron}, {Hatzidimitriou}, {Hauser}, {Haywood}, {Heiter}, {Helmi}, {Heu},
  {Hilger}, {Hobbs}, {Hofmann}, {Holland}, {Huckle}, {Hypki}, {Icardi},
  {Jan{\ss}en}, {Jevardat de Fombelle}, {Jonker}, {Juh{\'a}sz}, {Julbe},
  {Karampelas}, {Kewley}, {Klar}, {Kochoska}, {Kohley}, {Kolenberg},
  {Kontizas}, {Kontizas}, {Koposov}, {Kordopatis}, {Kostrzewa-Rutkowska},
  {Koubsky}, {Lambert}, {Lanza}, {Lasne}, {Lavigne}, {Le Fustec}, {Le
  Poncin-Lafitte}, {Lebreton}, {Leccia}, {Leclerc}, {Lecoeur-Taibi},
  {Lenhardt}, {Leroux}, {Liao}, {Licata}, {Lindstr{\o}m}, {Lister}, {Livanou},
  {Lobel}, {L{\'o}pez}, {Managau}, {Mann}, {Mantelet}, {Marchal}, {Marchant},
  {Marconi}, {Marinoni}, {Marschalk{\'o}}, {Marshall}, {Martino}, {Marton},
  {Mary}, {Massari}, {Matijevi{\v{c}}}, {Mazeh}, {McMillan}, {Messina},
  {Michalik}, {Millar}, {Molina}, {Molinaro}, {Moln{\'a}r}, {Montegriffo},
  {Mor}, {Morbidelli}, {Morel}, {Morris}, {Mulone}, {Muraveva}, {Musella},
  {Nelemans}, {Nicastro}, {Noval}, {O'Mullane}, {Ord{\'e}novic},
  {Ord{\'o}{\~n}ez-Blanco}, {Osborne}, {Pagani}, {Pagano}, {Pailler},
  {Palacin}, {Palaversa}, {Panahi}, {Pawlak}, {Piersimoni}, {Pineau}, {Plachy},
  {Plum}, {Poggio}, {Poujoulet}, {Pr{\v{s}}a}, {Pulone}, {Racero}, {Ragaini},
  {Rambaux}, {Ramos-Lerate}, {Regibo}, {Reyl{\'e}}, {Riclet}, {Ripepi}, {Riva},
  {Rivard}, {Rixon}, {Roegiers}, {Roelens}, {Romero-G{\'o}mez}, {Rowell},
  {Royer}, {Ruiz-Dern}, {Sadowski}, {Sagrist{\`a} Sell{\'e}s}, {Sahlmann},
  {Salgado}, {Salguero}, {Sanna}, {Santana-Ros}, {Sarasso}, {Savietto},
  {Schultheis}, {Sciacca}, {Segol}, {Segovia}, {S{\'e}gransan}, {Shih},
  {Siltala}, {Silva}, {Smart}, {Smith}, {Solano}, {Solitro}, {Sordo}, {Soria
  Nieto}, {Souchay}, {Spagna}, {Spoto}, {Stampa}, {Steele},
  {Steidelm{\"u}ller}, {Stephenson}, {Stoev}, {Suess}, {Surdej}, {Szabados},
  {Szegedi-Elek}, {Tapiador}, {Taris}, {Tauran}, {Taylor}, {Teixeira},
  {Terrett}, {Teyssandier}, {Thuillot}, {Titarenko}, {Torra Clotet}, {Turon},
  {Ulla}, {Utrilla}, {Uzzi}, {Vaillant}, {Valentini}, {Valette}, {van Elteren},
  {Van Hemelryck}, {van Leeuwen}, {Vaschetto}, {Vecchiato}, {Veljanoski},
  {Viala}, {Vicente}, {Vogt}, {von Essen}, {Voss}, {Votruba}, {Voutsinas},
  {Walmsley}, {Weiler}, {Wertz}, {Wevers}, {Wyrzykowski}, {Yoldas},
  {{\v{Z}}erjal}, {Ziaeepour}, {Zorec}, {Zschocke}, {Zucker}, {Zurbach}, \&
  {Zwitter}}]{GAIADR2}
{Gaia Collaboration}, {Brown}, A.~G.~A., {Vallenari}, A., {et~al.} 2018, \aap,
  616, A1

\bibitem[{{Gallagher} {et~al.}(2006){Gallagher}, {Brandt}, {Chartas},
  {Priddey}, {Garmire}, \& {Sambruna}}]{Gallagher06}
{Gallagher}, S.~C., {Brandt}, W.~N., {Chartas}, G., {et~al.} 2006, \apj, 644,
  709

\bibitem[{{Gallerani} {et~al.}(2017){Gallerani}, {Zappacosta}, {Orofino},
  {Piconcelli}, {Vignali}, {Ferrara}, {Maiolino}, {Fiore}, {Gilli},
  {Pallottini}, {Neri}, \& {Feruglio}}]{Gallerani17}
{Gallerani}, S., {Zappacosta}, L., {Orofino}, M.~C., {et~al.} 2017, \mnras,
  467, 3590

\bibitem[{{Gibson} {et~al.}(2009){Gibson}, {Jiang}, {Brandt}, {Hall}, {Shen},
  {Wu}, {Anderson}, {Schneider}, {Vanden Berk}, {Gallagher}, {Fan}, \&
  {York}}]{Gibson09a}
{Gibson}, R.~R., {Jiang}, L., {Brandt}, W.~N., {et~al.} 2009, \apj, 692, 758

\bibitem[{{Jiang} {et~al.}(2016){Jiang}, {McGreer}, {Fan}, {Strauss},
  {Ba{\~n}ados}, {Becker}, {Bian}, {Farnsworth}, {Shen}, {Wang}, {Wang},
  {Wang}, {White}, {Wu}, {Wu}, {Yang}, \& {Yang}}]{Jiang16}
{Jiang}, L., {McGreer}, I.~D., {Fan}, X., {et~al.} 2016, \apj, 833, 222

\bibitem[{{Joye} \& {Mandel}(2003)}]{Joye03}
{Joye}, W.~A. \& {Mandel}, E. 2003, in Astronomical Society of the Pacific
  Conference Series, Vol. 295, Astronomical Data Analysis Software and Systems
  XII, ed. H.~E. {Payne}, R.~I. {Jedrzejewski}, \& R.~N. {Hook}, 489

\bibitem[{{Just} {et~al.}(2007){Just}, {Brandt}, {Shemmer}, {Steffen},
  {Schneider}, {Chartas}, \& {Garmire}}]{Just07}
{Just}, D.~W., {Brandt}, W.~N., {Shemmer}, O., {et~al.} 2007, \apj, 665, 1004

\bibitem[{{Kalberla} {et~al.}(2005){Kalberla}, {Burton}, {Hartmann}, {Arnal},
  {Bajaja}, {Morras}, \& {P{\"o}ppel}}]{Kalberla05}
{Kalberla}, P.~M.~W., {Burton}, W.~B., {Hartmann}, D., {et~al.} 2005, \aap,
  440, 775

\bibitem[{{Leighly} {et~al.}(2007){Leighly}, {Halpern}, {Jenkins}, {Grupe},
  {Choi}, \& {Prescott}}]{Leighly07}
{Leighly}, K.~M., {Halpern}, J.~P., {Jenkins}, E.~B., {et~al.} 2007, \apj, 663,
  103

\bibitem[{{Liu} {et~al.}(2021){Liu}, {Luo}, {Brandt}, {Brotherton},
  {Gallagher}, {Ni}, {Shemmer}, \& {Timlin}}]{Liu21}
{Liu}, H., {Luo}, B., {Brandt}, W.~N., {et~al.} 2021, arXiv e-prints,
  arXiv:2102.02832

\bibitem[{{Liu} {et~al.}(2018){Liu}, {Luo}, {Brandt}, {Gallagher}, \&
  {Garmire}}]{Liu18}
{Liu}, H., {Luo}, B., {Brandt}, W.~N., {Gallagher}, S.~C., \& {Garmire}, G.~P.
  2018, \apj, 859, 113

\bibitem[{{Luo} {et~al.}(2013){Luo}, {Brandt}, {Alexander}, {Harrison},
  {Stern}, {Bauer}, {Boggs}, {Christensen}, {Comastri}, {Craig}, {Fabian},
  {Farrah}, {Fiore}, {Fuerst}, {Grefenstette}, {Hailey}, {Hickox}, {Madsen},
  {Matt}, {Ogle}, {Risaliti}, {Saez}, {Teng}, {Walton}, \& {Zhang}}]{Luo13}
{Luo}, B., {Brandt}, W.~N., {Alexander}, D.~M., {et~al.} 2013, \apj, 772, 153

\bibitem[{{Luo} {et~al.}(2014){Luo}, {Brandt}, {Alexander}, {Stern}, {Teng},
  {Ar{\'e}valo}, {Bauer}, {Boggs}, {Christensen}, {Comastri}, {Craig},
  {Farrah}, {Gandhi}, {Hailey}, {Harrison}, {Koss}, {Ogle}, {Puccetti}, {Saez},
  {Scott}, {Walton}, \& {Zhang}}]{Luo14}
{Luo}, B., {Brandt}, W.~N., {Alexander}, D.~M., {et~al.} 2014, \apj, 794, 70

\bibitem[{{Luo} {et~al.}(2015){Luo}, {Brandt}, {Hall}, {Wu}, {Anderson},
  {Garmire}, {Gibson}, {Plotkin}, {Richards}, {Schneider}, {Shemmer}, \&
  {Shen}}]{Luo15}
{Luo}, B., {Brandt}, W.~N., {Hall}, P.~B., {et~al.} 2015, \apj, 805, 122

\bibitem[{{Lusso} {et~al.}(2012){Lusso}, {Comastri}, {Simmons}, {Mignoli},
  {Zamorani}, {Vignali}, {Brusa}, {Shankar}, {Lutz}, {Trump}, {Maiolino},
  {Gilli}, {Bolzonella}, {Puccetti}, {Salvato}, {Impey}, {Civano}, {Elvis},
  {Mainieri}, {Silverman}, {Koekemoer}, {Bongiorno}, {Merloni}, {Berta}, {Le
  Floc'h}, {Magnelli}, {Pozzi}, \& {Riguccini}}]{Lusso12}
{Lusso}, E., {Comastri}, A., {Simmons}, B.~D., {et~al.} 2012, \mnras, 425, 623

\bibitem[{{Lusso} \& {Risaliti}(2016)}]{Lusso16}
{Lusso}, E. \& {Risaliti}, G. 2016, \apj, 819, 154

\bibitem[{{Martocchia} {et~al.}(2017){Martocchia}, {Piconcelli}, {Zappacosta},
  {Duras}, {Vietri}, {Vignali}, {Bianchi}, {Bischetti}, {Bongiorno}, {Brusa},
  {Lanzuisi}, {Marconi}, {Mathur}, {Miniutti}, {Nicastro}, {Bruni}, \&
  {Fiore}}]{Martocchia}
{Martocchia}, S., {Piconcelli}, E., {Zappacosta}, L., {et~al.} 2017, \aap, 608,
  A51

\bibitem[{{Matsuoka} {et~al.}(2018{\natexlab{a}}){Matsuoka}, {Onoue},
  {Kashikawa}, {Iwasawa}, {Strauss}, {Nagao}, {Imanishi}, {Lee}, {Akiyama},
  {Asami}, {Bosch}, {Foucaud}, {Furusawa}, {Goto}, {Gunn}, {Harikane}, {Ikeda},
  {Izumi}, {Kawaguchi}, {Kikuta}, {Kohno}, {Komiyama}, {Lupton}, {Minezaki},
  {Miyazaki}, {Morokuma}, {Murayama}, {Niida}, {Nishizawa}, {Oguri}, {Ono},
  {Ouchi}, {Price}, {Sameshima}, {Schulze}, {Shirakata}, {Silverman},
  {Sugiyama}, {Tait}, {Takada}, {Takata}, {Tanaka}, {Tang}, {Toba}, {Utsumi},
  \& {Wang}}]{Matsuoka18a}
{Matsuoka}, Y., {Onoue}, M., {Kashikawa}, N., {et~al.} 2018{\natexlab{a}},
  \pasj, 70, S35

\bibitem[{{Matsuoka} {et~al.}(2016){Matsuoka}, {Onoue}, {Kashikawa}, {Iwasawa},
  {Strauss}, {Nagao}, {Imanishi}, {Niida}, {Toba}, {Akiyama}, {Asami}, {Bosch},
  {Foucaud}, {Furusawa}, {Goto}, {Gunn}, {Harikane}, {Ikeda}, {Kawaguchi},
  {Kikuta}, {Komiyama}, {Lupton}, {Minezaki}, {Miyazaki}, {Morokuma},
  {Murayama}, {Nishizawa}, {Ono}, {Ouchi}, {Price}, {Sameshima}, {Silverman},
  {Sugiyama}, {Tait}, {Takada}, {Takata}, {Tanaka}, {Tang}, \&
  {Utsumi}}]{Matsuoka16}
{Matsuoka}, Y., {Onoue}, M., {Kashikawa}, N., {et~al.} 2016, \apj, 828, 26

\bibitem[{{Matsuoka} {et~al.}(2019){Matsuoka}, {Onoue}, {Kashikawa}, {Strauss},
  {Iwasawa}, {Lee}, {Imanishi}, {Nagao}, {Akiyama}, {Asami}, {Bosch},
  {Furusawa}, {Goto}, {Gunn}, {Harikane}, {Ikeda}, {Izumi}, {Kawaguchi},
  {Kato}, {Kikuta}, {Kohno}, {Komiyama}, {Koyama}, {Lupton}, {Minezaki},
  {Miyazaki}, {Murayama}, {Niida}, {Nishizawa}, {Noboriguchi}, {Oguri}, {Ono},
  {Ouchi}, {Price}, {Sameshima}, {Schulze}, {Shirakata}, {Silverman},
  {Sugiyama}, {Tait}, {Takada}, {Takata}, {Tanaka}, {Tang}, {Toba}, {Utsumi},
  {Wang}, \& {Yamashita}}]{Matsuoka19}
{Matsuoka}, Y., {Onoue}, M., {Kashikawa}, N., {et~al.} 2019, \apj, 872, L2

\bibitem[{{Matsuoka} {et~al.}(2018{\natexlab{b}}){Matsuoka}, {Strauss},
  {Kashikawa}, {Onoue}, {Iwasawa}, {Tang}, {Lee}, {Imanishi}, {Nagao},
  {Akiyama}, {Asami}, {Bosch}, {Furusawa}, {Goto}, {Gunn}, {Harikane}, {Ikeda},
  {Izumi}, {Kawaguchi}, {Kato}, {Kikuta}, {Kohno}, {Komiyama}, {Lupton},
  {Minezaki}, {Miyazaki}, {Murayama}, {Niida}, {Nishizawa}, {Noboriguchi},
  {Oguri}, {Ono}, {Ouchi}, {Price}, {Sameshima}, {Schulze}, {Shirakata},
  {Silverman}, {Sugiyama}, {Tait}, {Takada}, {Takata}, {Tanaka}, {Toba},
  {Utsumi}, {Wang}, \& {Yamashita}}]{Matsuoka18c}
{Matsuoka}, Y., {Strauss}, M.~A., {Kashikawa}, N., {et~al.} 2018{\natexlab{b}},
  \apj, 869, 150

\bibitem[{{Mazzucchelli} {et~al.}(2017){Mazzucchelli}, {Ba{\~n}ados},
  {Venemans}, {Decarli}, {Farina}, {Walter}, {Eilers}, {Rix}, {Simcoe},
  {Stern}, {Fan}, {Schlafly}, {De Rosa}, {Hennawi}, {Chambers}, {Greiner},
  {Burgett}, {Draper}, {Kaiser}, {Kudritzki}, {Magnier}, {Metcalfe}, {Waters},
  \& {Wainscoat}}]{Mazzucchelli17b}
{Mazzucchelli}, C., {Ba{\~n}ados}, E., {Venemans}, B.~P., {et~al.} 2017, \apj,
  849, 91

\bibitem[{{Mazzucchelli} {et~al.}(2019){Mazzucchelli}, {Decarli}, {Farina},
  {Ba{\~n}ados}, {Venemans}, {Strauss}, {Walter}, {Neeleman}, {Bertoldi},
  {Fan}, {Riechers}, {Rix}, \& {Wang}}]{Mazzucchelli19}
{Mazzucchelli}, C., {Decarli}, R., {Farina}, E.~P., {et~al.} 2019, \apj, 881,
  163

\bibitem[{{Meyer} {et~al.}(2019){Meyer}, {Bosman}, \& {Ellis}}]{Meyer19}
{Meyer}, R.~A., {Bosman}, S. E.~I., \& {Ellis}, R.~S. 2019, \mnras, 487, 3305

\bibitem[{{Mignoli} {et~al.}(2020){Mignoli}, {Gilli}, {Decarli}, {Vanzella},
  {Balmaverde}, {Cappelluti}, {Cassar{\`a}}, {Comastri}, {Cusano}, {Iwasawa},
  {Marchesi}, {Prandoni}, {Vignali}, {Vito}, {Zamorani}, {Chiaberge}, \&
  {Norman}}]{Mignoli20}
{Mignoli}, M., {Gilli}, R., {Decarli}, R., {et~al.} 2020, \aap, 642, L1

\bibitem[{{Murphy} \& {Yaqoob}(2009)}]{Murphy09}
{Murphy}, K.~D. \& {Yaqoob}, T. 2009, \mnras, 397, 1549

\bibitem[{{Nanni} {et~al.}(2018){Nanni}, {Gilli}, {Vignali}, {Mignoli},
  {Comastri}, {Vanzella}, {Zamorani}, {Calura}, {Lanzuisi}, {Brusa}, {Tozzi},
  {Iwasawa}, {Cappi}, {Vito}, {Balmaverde}, {Costa}, {Risaliti}, {Paolillo},
  {Prandoni}, {Liuzzo}, {Rosati}, {Chiaberge}, {Caminha}, {Sani}, {Cappelluti},
  \& {Norman}}]{Nanni18}
{Nanni}, R., {Gilli}, R., {Vignali}, C., {et~al.} 2018, \aap, 614, A121

\bibitem[{{Nanni} {et~al.}(2017){Nanni}, {Vignali}, {Gilli}, {Moretti}, \&
  {Brandt}}]{Nanni17}
{Nanni}, R., {Vignali}, C., {Gilli}, R., {Moretti}, A., \& {Brandt}, W.~N.
  2017, \aap, 603, A128

\bibitem[{{Neeleman} {et~al.}(2019){Neeleman}, {Ba{\~n}ados}, {Walter},
  {Decarli}, {Venemans}, {Carilli}, {Fan}, {Farina}, {Mazzucchelli}, {Novak},
  {Riechers}, {Rix}, \& {Wang}}]{Neeleman19}
{Neeleman}, M., {Ba{\~n}ados}, E., {Walter}, F., {et~al.} 2019, \apj, 882, 10

\bibitem[{{Ni} {et~al.}(2018){Ni}, {Brandt}, {Luo}, {Hall}, {Shen}, {Anderson},
  {Plotkin}, {Richards}, {Schneider}, {Shemmer}, \& {Wu}}]{Ni18}
{Ni}, Q., {Brandt}, W.~N., {Luo}, B., {et~al.} 2018, \mnras, 480, 5184

\bibitem[{{Onoue} {et~al.}(2020){Onoue}, {Ba{\~n}ados}, {Mazzucchelli},
  {Venemans}, {Schindler}, {Walter}, {Hennawi}, {Andika}, {Davies}, {Decarli},
  {Farina}, {Jahnke}, {Nagao}, {Tominaga}, \& {Wang}}]{Onoue20}
{Onoue}, M., {Ba{\~n}ados}, E., {Mazzucchelli}, C., {et~al.} 2020, \apj, 898,
  105

\bibitem[{{Onoue} {et~al.}(2019){Onoue}, {Kashikawa}, {Matsuoka}, {Kato},
  {Izumi}, {Nagao}, {Strauss}, {Harikane}, {Imanishi}, {Ito}, {Iwasawa},
  {Kawaguchi}, {Lee}, {Noboriguchi}, {Suh}, {Tanaka}, \& {Toba}}]{Onoue19}
{Onoue}, M., {Kashikawa}, N., {Matsuoka}, Y., {et~al.} 2019, \apj, 880, 77

\bibitem[{{Pacucci} {et~al.}(2015){Pacucci}, {Ferrara}, {Volonteri}, \&
  {Dubus}}]{Pacucci15}
{Pacucci}, F., {Ferrara}, A., {Volonteri}, M., \& {Dubus}, G. 2015, \mnras,
  454, 3771

\bibitem[{{Paolillo} {et~al.}(2017){Paolillo}, {Papadakis}, {Brandt}, {Luo},
  {Xue}, {Tozzi}, {Shemmer}, {Allevato}, {Bauer}, {Comastri}, {Gilli},
  {Koekemoer}, {Liu}, {Vignali}, {Vito}, {Yang}, {Wang}, \&
  {Zheng}}]{Paolillo17}
{Paolillo}, M., {Papadakis}, I., {Brandt}, W.~N., {et~al.} 2017, \mnras, 471,
  4398

\bibitem[{{Planck Collaboration} {et~al.}(2016){Planck Collaboration}, {Ade},
  {Aghanim}, {Arnaud}, {Ashdown}, {Aumont}, {Baccigalupi}, {Banday},
  {Barreiro}, {Bartlett}, {Bartolo}, {Battaner}, {Battye}, {Benabed},
  {Beno{\^\i}t}, {Benoit-L{\'e}vy}, {Bernard}, {Bersanelli}, {Bielewicz},
  {Bock}, {Bonaldi}, {Bonavera}, {Bond}, {Borrill}, {Bouchet}, {Boulanger},
  {Bucher}, {Burigana}, {Butler}, {Calabrese}, {Cardoso}, {Catalano},
  {Challinor}, {Chamballu}, {Chary}, {Chiang}, {Chluba}, {Christensen},
  {Church}, {Clements}, {Colombi}, {Colombo}, {Combet}, {Coulais}, {Crill},
  {Curto}, {Cuttaia}, {Danese}, {Davies}, {Davis}, {de Bernardis}, {de Rosa},
  {de Zotti}, {Delabrouille}, {D{\'e}sert}, {Di Valentino}, {Dickinson},
  {Diego}, {Dolag}, {Dole}, {Donzelli}, {Dor{\'e}}, {Douspis}, {Ducout},
  {Dunkley}, {Dupac}, {Efstathiou}, {Elsner}, {En{\ss}lin}, {Eriksen},
  {Farhang}, {Fergusson}, {Finelli}, {Forni}, {Frailis}, {Fraisse},
  {Franceschi}, {Frejsel}, {Galeotta}, {Galli}, {Ganga}, {Gauthier}, {Gerbino},
  {Ghosh}, {Giard}, {Giraud-H{\'e}raud}, {Giusarma}, {Gjerl{\o}w},
  {Gonz{\'a}lez-Nuevo}, {G{\'o}rski}, {Gratton}, {Gregorio}, {Gruppuso},
  {Gudmundsson}, {Hamann}, {Hansen}, {Hanson}, {Harrison}, {Helou}, {Henrot-
  Versill{\'e}}, {Hern{\'a}ndez-Monteagudo}, {Herranz}, {Hildebrandt}, {Hivon},
  {Hobson}, {Holmes}, {Hornstrup}, {Hovest}, {Huang}, {Huffenberger}, {Hurier},
  {Jaffe}, {Jaffe}, {Jones}, {Juvela}, {Keih{\"a}nen}, {Keskitalo}, {Kisner},
  {Kneissl}, {Knoche}, {Knox}, {Kunz}, {Kurki-Suonio}, {Lagache},
  {L{\"a}hteenm{\"a}ki}, {Lamarre}, {Lasenby}, {Lattanzi}, {Lawrence}, {Leahy},
  {Leonardi}, {Lesgourgues}, {Levrier}, {Lewis}, {Liguori}, {Lilje},
  {Linden-V{\o}rnle}, {L{\'o}pez-Caniego}, {Lubin}, {Mac{\'\i}as-P{\'e}rez},
  {Maggio}, {Maino}, {Mandolesi}, {Mangilli}, {Marchini}, {Maris}, {Martin},
  {Martinelli}, {Mart{\'\i}nez-Gonz{\'a}lez}, {Masi}, {Matarrese}, {McGehee},
  {Meinhold}, {Melchiorri}, {Melin}, {Mendes}, {Mennella}, {Migliaccio},
  {Millea}, {Mitra}, {Miville-Desch{\^e}nes}, {Moneti}, {Montier}, {Morgante},
  {Mortlock}, {Moss}, {Munshi}, {Murphy}, {Naselsky}, {Nati}, {Natoli},
  {Netterfield}, {N{\o}rgaard-Nielsen}, {Noviello}, {Novikov}, {Novikov},
  {Oxborrow}, {Paci}, {Pagano}, {Pajot}, {Paladini}, {Paoletti}, {Partridge},
  {Pasian}, {Patanchon}, {Pearson}, {Perdereau}, {Perotto}, {Perrotta},
  {Pettorino}, {Piacentini}, {Piat}, {Pierpaoli}, {Pietrobon}, {Plaszczynski},
  {Pointecouteau}, {Polenta}, {Popa}, {Pratt}, {Pr{\'e}zeau}, {Prunet},
  {Puget}, {Rachen}, {Reach}, {Rebolo}, {Reinecke}, {Remazeilles}, {Renault},
  {Renzi}, {Ristorcelli}, {Rocha}, {Rosset}, {Rossetti}, {Roudier},
  {Rouill{\'e} d'Orfeuil}, {Rowan-Robinson}, {Rubi{\~n}o-Mart{\'\i}n},
  {Rusholme}, {Said}, {Salvatelli}, {Salvati}, {Sandri}, {Santos},
  {Savelainen}, {Savini}, {Scott}, {Seiffert}, {Serra}, {Shellard}, {Spencer},
  {Spinelli}, {Stolyarov}, {Stompor}, {Sudiwala}, {Sunyaev}, {Sutton},
  {Suur-Uski}, {Sygnet}, {Tauber}, {Terenzi}, {Toffolatti}, {Tomasi},
  {Tristram}, {Trombetti}, {Tucci}, {Tuovinen}, {T{\"u}rler}, {Umana},
  {Valenziano}, {Valiviita}, {Van Tent}, {Vielva}, {Villa}, {Wade}, {Wandelt},
  {Wehus}, {White}, {White}, {Wilkinson}, {Yvon}, {Zacchei}, \&
  {Zonca}}]{Planck16}
{Planck Collaboration}, {Ade}, P.~A.~R., {Aghanim}, N., {et~al.} 2016, \aap,
  594

\bibitem[{{Plotkin} {et~al.}(2015){Plotkin}, {Shemmer}, {Trakhtenbrot},
  {Anderson}, {Brandt}, {Fan}, {Gallo}, {Lira}, {Luo}, {Richards}, {Schneider},
  {Strauss}, \& {Wu}}]{Plotkin15}
{Plotkin}, R.~M., {Shemmer}, O., {Trakhtenbrot}, B., {et~al.} 2015, \apj, 805,
  123

\bibitem[{{Pons} {et~al.}(2020){Pons}, {McMahon}, {Banerji}, \&
  {Reed}}]{Pons20}
{Pons}, E., {McMahon}, R.~G., {Banerji}, M., \& {Reed}, S.~L. 2020, \mnras,
  491, 3884

\bibitem[{{Pons} {et~al.}(2021){Pons}, {McMahon}, {Banerji}, \&
  {Reed}}]{Pons21}
{Pons}, E., {McMahon}, R.~G., {Banerji}, M., \& {Reed}, S.~L. 2021, \mnras,
  501, 6208

\bibitem[{{Proga} \& {Kallman}(2004)}]{Proga04}
{Proga}, D. \& {Kallman}, T.~R. 2004, \apj, 616, 688

\bibitem[{{Proga} {et~al.}(2000){Proga}, {Stone}, \& {Kallman}}]{Proga00}
{Proga}, D., {Stone}, J.~M., \& {Kallman}, T.~R. 2000, \apj, 543, 686

\bibitem[{{Pu} {et~al.}(2020){Pu}, {Luo}, {Brandt}, {Timlin}, {Liu}, {Ni}, \&
  {Wu}}]{Pu20}
{Pu}, X., {Luo}, B., {Brandt}, W.~N., {et~al.} 2020, \apj, 900, 141

\bibitem[{{Reed} {et~al.}(2019){Reed}, {Banerji}, {Becker}, {Hewett},
  {Martini}, {McMahon}, {Pons}, {Rauch}, {Abbott}, \& {Allam}}]{Reed19}
{Reed}, S.~L., {Banerji}, M., {Becker}, G.~D., {et~al.} 2019, \mnras, 487, 1874

\bibitem[{{Richards} {et~al.}(2011){Richards}, {Kruczek}, {Gallagher}, {Hall},
  {Hewett}, {Leighly}, {Deo}, {Kratzer}, \& {Shen}}]{Richards11}
{Richards}, G.~T., {Kruczek}, N.~E., {Gallagher}, S.~C., {et~al.} 2011, \aj,
  141, 167

\bibitem[{{Risaliti} {et~al.}(2011){Risaliti}, {Nardini}, {Salvati}, {Elvis},
  {Fabbiano}, {Maiolino}, {Pietrini}, \& {Torricelli-Ciamponi}}]{Risaliti11}
{Risaliti}, G., {Nardini}, E., {Salvati}, M., {et~al.} 2011, \mnras, 410, 1027

\bibitem[{{Salvestrini} {et~al.}(2019){Salvestrini}, {Risaliti}, {Bisogni},
  {Lusso}, \& {Vignali}}]{Salvestrini19}
{Salvestrini}, F., {Risaliti}, G., {Bisogni}, S., {Lusso}, E., \& {Vignali}, C.
  2019, \aap, 631, A120

\bibitem[{{Schindler} {et~al.}(2020){Schindler}, {Farina}, {Ba{\~n}ados},
  {Eilers}, {Hennawi}, {Onoue}, {Venemans}, {Walter}, {Wang}, {Davies},
  {Decarli}, {Rosa}, {Drake}, {Fan}, {Mazzucchelli}, {Rix}, {Worseck}, \&
  {Yang}}]{Schindler20}
{Schindler}, J.-T., {Farina}, E.~P., {Ba{\~n}ados}, E., {et~al.} 2020, \apj,
  905, 51

\bibitem[{{Schlafly} \& {Finkbeiner}(2011)}]{Schlafly11}
{Schlafly}, E.~F. \& {Finkbeiner}, D.~P. 2011, \apj, 737, 103

\bibitem[{{Selsing} {et~al.}(2016){Selsing}, {Fynbo}, {Christensen}, \&
  {Krogager}}]{Selsing16}
{Selsing}, J., {Fynbo}, J.~P.~U., {Christensen}, L., \& {Krogager}, J.~K. 2016,
  \aap, 585, A87

\bibitem[{{Shemmer} {et~al.}(2006){Shemmer}, {Brandt}, {Netzer}, {Maiolino}, \&
  {Kaspi}}]{Shemmer06}
{Shemmer}, O., {Brandt}, W.~N., {Netzer}, H., {Maiolino}, R., \& {Kaspi}, S.
  2006, \apjl, 646, L29

\bibitem[{{Shemmer} {et~al.}(2008){Shemmer}, {Brandt}, {Netzer}, {Maiolino}, \&
  {Kaspi}}]{Shemmer08}
{Shemmer}, O., {Brandt}, W.~N., {Netzer}, H., {Maiolino}, R., \& {Kaspi}, S.
  2008, \apj, 682, 81

\bibitem[{{Shemmer} {et~al.}(2017){Shemmer}, {Brandt}, {Paolillo}, {Kaspi},
  {Vignali}, {Lira}, \& {Schneider}}]{Shemmer17}
{Shemmer}, O., {Brandt}, W.~N., {Paolillo}, M., {et~al.} 2017, \apj, 848, 46

\bibitem[{{Shemmer} {et~al.}(2005){Shemmer}, {Brandt}, {Vignali}, {Schneider},
  {Fan}, {Richards}, \& {Strauss}}]{Shemmer05}
{Shemmer}, O., {Brandt}, W.~N., {Vignali}, C., {et~al.} 2005, \apj, 630, 729

\bibitem[{{Shen}(2013)}]{Shen13}
{Shen}, Y. 2013, Bulletin of the Astronomical Society of India, 41, 61

\bibitem[{{Shen} {et~al.}(2011){Shen}, {Richards}, {Strauss}, {Hall},
  {Schneider}, {Snedden}, {Bizyaev}, {Brewington}, {Malanushenko},
  {Malanushenko}, {Oravetz}, {Pan}, \& {Simmons}}]{Shen11}
{Shen}, Y., {Richards}, G.~T., {Strauss}, M.~A., {et~al.} 2011, \apjs, 194, 45

\bibitem[{{Shen} {et~al.}(2019){Shen}, {Wu}, {Jiang}, {Ba{\~n}ados}, {Fan},
  {Ho}, {Riechers}, {Strauss}, {Venemans}, {Vestergaard}, {Walter}, {Wang},
  {Willott}, {Wu}, \& {Yang}}]{Shen19}
{Shen}, Y., {Wu}, J., {Jiang}, L., {et~al.} 2019, \apj, 873, 35

\bibitem[{{Simcoe} {et~al.}(2008){Simcoe}, {Burgasser}, {Bernstein}, {Bigelow},
  {Fishner}, {Forrest}, {McMurtry}, {Pipher}, {Schechter}, \&
  {Smith}}]{Simcoe08}
{Simcoe}, R.~A., {Burgasser}, A.~J., {Bernstein}, R.~A., {et~al.} 2008, in
  Society of Photo-Optical Instrumentation Engineers (SPIE) Conference Series,
  Vol. 7014, Ground-based and Airborne Instrumentation for Astronomy II, ed.
  I.~S. {McLean} \& M.~M. {Casali}, 70140U

\bibitem[{{Spingola} {et~al.}(2020){Spingola}, {Dallacasa}, {Belladitta},
  {Caccianiga}, {Giroletti}, {Moretti}, \& {Orienti}}]{Spingola20}
{Spingola}, C., {Dallacasa}, D., {Belladitta}, S., {et~al.} 2020, \aap, 643,
  L12

\bibitem[{{Steffen} {et~al.}(2006){Steffen}, {Strateva}, {Brandt}, {Alexander},
  {Koekemoer}, {Lehmer}, {Schneider}, \& {Vignali}}]{Steffen06}
{Steffen}, A.~T., {Strateva}, I., {Brandt}, W.~N., {et~al.} 2006, \aj, 131,
  2826

\bibitem[{{Stoehr} {et~al.}(2008){Stoehr}, {White}, {Smith}, {Kamp},
  {Thompson}, {Durand}, {Freudling}, {Fraquelli}, {Haase}, {Hook}, {Kimball},
  {K{\"u}mmel}, {Levay}, {Lombardi}, {Micol}, \& {Rogers}}]{Stoehr08}
{Stoehr}, F., {White}, R., {Smith}, M., {et~al.} 2008, in Astronomical Society
  of the Pacific Conference Series, Vol. 394, Astronomical Data Analysis
  Software and Systems XVII, ed. R.~W. {Argyle}, P.~S. {Bunclark}, \& J.~R.
  {Lewis}, 505

\bibitem[{{Timlin} {et~al.}(2020{\natexlab{a}}){Timlin}, {Brandt}, {Zhu},
  {Liu}, {Luo}, \& {Ni}}]{Timlin20b}
{Timlin}, John~D., I., {Brandt}, W.~N., {Zhu}, S., {et~al.} 2020{\natexlab{a}},
  \mnras, 498, 4033

\bibitem[{{Timlin} {et~al.}(2020{\natexlab{b}}){Timlin}, {Brandt}, {Ni}, {Luo},
  {Pu}, {Schneider}, {Vivek}, \& {Yi}}]{Timlin20}
{Timlin}, J.~D., {Brandt}, W.~N., {Ni}, Q., {et~al.} 2020{\natexlab{b}},
  \mnras, 492, 719

\bibitem[{{Valiante} {et~al.}(2017){Valiante}, {Agarwal}, {Habouzit}, \&
  {Pezzulli}}]{Valiante17}
{Valiante}, R., {Agarwal}, B., {Habouzit}, M., \& {Pezzulli}, E. 2017,
  Publications of the Astronomical Society of Australia, 34, e031

\bibitem[{{Vanden Berk} {et~al.}(2001){Vanden Berk}, {Richards}, {Bauer},
  {Strauss}, {Schneider}, {Heckman}, {York}, {Hall}, {Fan}, {Knapp},
  {Anderson}, {Annis}, {Bahcall}, {Bernardi}, {Briggs}, {Brinkmann}, {Brunner},
  {Burles}, {Carey}, {Castander}, {Connolly}, {Crocker}, {Csabai}, {Doi},
  {Finkbeiner}, {Friedman}, {Frieman}, {Fukugita}, {Gunn}, {Hennessy},
  {Ivezi{\'c}}, {Kent}, {Kunszt}, {Lamb}, {Leger}, {Long}, {Loveday}, {Lupton},
  {Meiksin}, {Merelli}, {Munn}, {Newberg}, {Newcomb}, {Nichol}, {Owen}, {Pier},
  {Pope}, {Rockosi}, {Schlegel}, {Siegmund}, {Smee}, {Snir}, {Stoughton},
  {Stubbs}, {SubbaRao}, {Szalay}, {Szokoly}, {Tremonti}, {Uomoto}, {Waddell},
  {Yanny}, \& {Zheng}}]{VandenBerk01}
{Vanden Berk}, D.~E., {Richards}, G.~T., {Bauer}, A., {et~al.} 2001, \aj, 122,
  549

\bibitem[{{Venemans} {et~al.}(2015){Venemans}, {Ba{\~n}ados}, {Decarli},
  {Farina}, {Walter}, {Chambers}, {Fan}, {Rix}, {Schlafly}, {McMahon},
  {Simcoe}, {Stern}, {Burgett}, {Draper}, {Flewelling}, {Hodapp}, {Kaiser},
  {Magnier}, {Metcalfe}, {Morgan}, {Price}, {Tonry}, {Waters}, {AlSayyad},
  {Banerji}, {Chen}, {Gonz{\'a}lez-Solares}, {Greiner}, {Mazzucchelli},
  {McGreer}, {Miller}, {Reed}, \& {Sullivan}}]{Venemans15a}
{Venemans}, B.~P., {Ba{\~n}ados}, E., {Decarli}, R., {et~al.} 2015, \apjl, 801,
  L11

\bibitem[{{Venemans} {et~al.}(2020){Venemans}, {Walter}, {Neeleman}, {Novak},
  {Otter}, {Decarli}, {Ba{\~n}ados}, {Drake}, {Farina}, {Kaasinen},
  {Mazzucchelli}, {Carilli}, {Fan}, {Rix}, \& {Wang}}]{Venemans20}
{Venemans}, B.~P., {Walter}, F., {Neeleman}, M., {et~al.} 2020, \apj, 904, 130

\bibitem[{{Venemans} {et~al.}(2016){Venemans}, {Walter}, {Zschaechner},
  {Decarli}, {De Rosa}, {Findlay}, {McMahon}, \& {Sutherland}}]{Venemans16}
{Venemans}, B.~P., {Walter}, F., {Zschaechner}, L., {et~al.} 2016, \apj, 816,
  37

\bibitem[{{Vestergaard} \& {Osmer}(2009)}]{Vestergaard09}
{Vestergaard}, M. \& {Osmer}, P.~S. 2009, \apj, 699, 800

\bibitem[{{Vestergaard} \& {Wilkes}(2001)}]{Vestergaard01}
{Vestergaard}, M. \& {Wilkes}, B.~J. 2001, \apjs, 134, 1

\bibitem[{{Vietri} {et~al.}(2020){Vietri}, {Mainieri}, {Kakkad}, {Netzer},
  {Perna}, {Circosta}, {Harrison}, {Zappacosta}, {Husemann}, {Padovani},
  {Bischetti}, {Bongiorno}, {Brusa}, {Carniani}, {Cicone}, {Comastri},
  {Cresci}, {Feruglio}, {Fiore}, {Lanzuisi}, {Mannucci}, {Marconi},
  {Piconcelli}, {Puglisi}, {Salvato}, {Schramm}, {Schulze}, {Scholtz},
  {Vignali}, \& {Zamorani}}]{Vietri20}
{Vietri}, G., {Mainieri}, V., {Kakkad}, D., {et~al.} 2020, \aap, 644, A175

\bibitem[{{Vietri} {et~al.}(2018){Vietri}, {Piconcelli}, {Bischetti}, {Duras},
  {Martocchia}, {Bongiorno}, {Marconi}, {Zappacosta}, {Bisogni}, {Bruni},
  {Brusa}, {Comastri}, {Cresci}, {Feruglio}, {Giallongo}, {La Franca},
  {Mainieri}, {Mannucci}, {Ricci}, {Sani}, {Testa}, {Tombesi}, {Vignali}, \&
  {Fiore}}]{Vietri18}
{Vietri}, G., {Piconcelli}, E., {Bischetti}, M., {et~al.} 2018, \aap, 617, A81

\bibitem[{{Vito} {et~al.}(2019{\natexlab{a}}){Vito}, {Brandt}, {Bauer},
  {Calura}, {Gilli}, {Luo}, {Shemmer}, {Vignali}, {Zamorani}, {Brusa},
  {Civano}, {Comastri}, \& {Nanni}}]{Vito19b}
{Vito}, F., {Brandt}, W.~N., {Bauer}, F.~E., {et~al.} 2019{\natexlab{a}}, \aap,
  630, A118

\bibitem[{{Vito} {et~al.}(2019{\natexlab{b}}){Vito}, {Brandt}, {Bauer},
  {Gilli}, {Luo}, {Zamorani}, {Calura}, {Comastri}, {Mazzucchelli}, {Mignoli},
  {Nanni}, {Shemmer}, {Vignali}, {Brusa}, {Cappelluti}, {Civano}, \&
  {Volonteri}}]{Vito19a}
{Vito}, F., {Brandt}, W.~N., {Bauer}, F.~E., {et~al.} 2019{\natexlab{b}}, \aap,
  628, L6

\bibitem[{{Vito} {et~al.}(2018{\natexlab{a}}){Vito}, {Brandt}, {Luo},
  {Shemmer}, {Vignali}, \& {Gilli}}]{Vito18c}
{Vito}, F., {Brandt}, W.~N., {Luo}, B., {et~al.} 2018{\natexlab{a}}, \mnras,
  479, 5335

\bibitem[{{Vito} {et~al.}(2018{\natexlab{b}}){Vito}, {Brandt}, {Stern},
  {Assef}, {Chen}, {Brightman}, {Comastri}, {Eisenhardt}, {Garmire}, {Hickox},
  {Lansbury}, {Tsai}, {Walton}, \& {Wu}}]{Vito18b}
{Vito}, F., {Brandt}, W.~N., {Stern}, D., {et~al.} 2018{\natexlab{b}}, \mnras,
  474, 4528

\bibitem[{{Wang} {et~al.}(2021{\natexlab{a}}){Wang}, {Fan}, {Yang},
  {Mazzucchelli}, {Wu}, {Li}, {Ba{\~n}ados}, {Farina}, {Nanni}, {Ai}, {Bian},
  {Davies}, {Decarli}, {Hennawi}, {Schindler}, {Venemans}, \&
  {Walter}}]{Wang21a}
{Wang}, F., {Fan}, X., {Yang}, J., {et~al.} 2021{\natexlab{a}}, \apj, 908, 53

\bibitem[{{Wang} {et~al.}(2021{\natexlab{b}}){Wang}, {Yang}, {Fan}, {Hennawi},
  {Barth}, {Banados}, {Bian}, {Boutsia}, {Connor}, {Davies}, {Decarli},
  {Eilers}, {Farina}, {Green}, {Jiang}, {Li}, {Mazzucchelli}, {Nanni},
  {Schindler}, {Venemans}, {Walter}, {Wu}, \& {Yue}}]{Wang21b}
{Wang}, F., {Yang}, J., {Fan}, X., {et~al.} 2021{\natexlab{b}}, \apjl, 907, L1

\bibitem[{{Weisskopf} {et~al.}(2007){Weisskopf}, {Wu}, {Trimble}, {O'Dell},
  {Elsner}, {Zavlin}, \& {Kouveliotou}}]{Weisskopf07}
{Weisskopf}, M.~C., {Wu}, K., {Trimble}, V., {et~al.} 2007, \apj, 657, 1026

\bibitem[{{Willott} {et~al.}(2017){Willott}, {Bergeron}, \&
  {Omont}}]{Willott17}
{Willott}, C.~J., {Bergeron}, J., \& {Omont}, A. 2017, \apj, 850, 108

\bibitem[{{Woods} {et~al.}(2019){Woods}, {Agarwal}, {Bromm}, {Bunker}, {Chen},
  {Chon}, {Ferrara}, {Glover}, {Haemmerl{\'e}}, {Haiman}, {Hartwig}, {Heger},
  {Hirano}, {Hosokawa}, {Inayoshi}, {Klessen}, {Kobayashi}, {Koliopanos},
  {Latif}, {Li}, {Mayer}, {Mezcua}, {Natarajan}, {Pacucci}, {Rees}, {Regan},
  {Sakurai}, {Salvadori}, {Schneider}, {Surace}, {Tanaka}, {Whalen}, \&
  {Yoshida}}]{Woods19}
{Woods}, T.~E., {Agarwal}, B., {Bromm}, V., {et~al.} 2019, \pasa, 36, e027

\bibitem[{{Wu} {et~al.}(2015){Wu}, {Wang}, {Fan}, {Yi}, {Zuo}, {Bian}, {Jiang},
  {McGreer}, {Wang}, {Yang}, {Yang}, {Thompson}, \& {Beletsky}}]{Wu15}
{Wu}, X.-B., {Wang}, F., {Fan}, X., {et~al.} 2015, \nat, 518, 512

\bibitem[{{Yang} {et~al.}(2020){Yang}, {Wang}, {Fan}, {Hennawi}, {Davies},
  {Yue}, {Banados}, {Wu}, {Venemans}, {Barth}, {Bian}, {Boutsia}, {Decarli},
  {Farina}, {Green}, {Jiang}, {Li}, {Mazzucchelli}, \& {Walter}}]{Yang20}
{Yang}, J., {Wang}, F., {Fan}, X., {et~al.} 2020, \apjl, 897, L14

\bibitem[{{Yi} {et~al.}(2019){Yi}, {Vivek}, {Brandt}, {Wang}, {Timlin}, {Filiz
  Ak}, {Schneider}, {Fynbo}, {Ni}, {Vito}, {Indahl}, \& {Sameer}}]{Yi19}
{Yi}, W., {Vivek}, M., {Brandt}, W.~N., {et~al.} 2019, \apj, 870, L25

\bibitem[{{Zappacosta} {et~al.}(2020){Zappacosta}, {Piconcelli}, {Giustini},
  {Vietri}, {Duras}, {Miniutti}, {Bischetti}, {Bongiorno}, {Brusa},
  {Chiaberge}, {Comastri}, {Feruglio}, {Luminari}, {Marconi}, {Ricci},
  {Vignali}, \& {Fiore}}]{Zappacosta20}
{Zappacosta}, L., {Piconcelli}, E., {Giustini}, M., {et~al.} 2020, \aap, 635,
  L5

\end{thebibliography}
%\begin{thebibliography}{}
%
%\end{thebibliography}
%
\end{document}